\documentclass[fleqn,10pt]{wlscirep}
\usepackage[utf8]{inputenc}
\usepackage[T1]{fontenc}

\usepackage{caption}
\usepackage{amssymb}
\usepackage{epsfig}
\usepackage{multirow}
\usepackage{amsmath}
\usepackage{times}
\usepackage{graphicx}
\usepackage{xcolor}
\usepackage{hyperref}
\hypersetup{
	colorlinks=true,
	citecolor=blue,
	linkcolor=black,
	urlcolor=blue
}

\usepackage{color,soul}

\title{Emergence of global synchronization in directed excitatory networks of type I neurons}

\author[1]{Abolfazl Ziaeemehr}
\author[1,2,*]{Mina Zarei}
\author[3]{Aida Sheshbolouki}
\affil[1]{Institute of Advanced Studies in Basic Sciences (IASBS),Department of Physics, Zanjan, 45137-66731, Iran.}
\affil[2]{Institute for Research in Fundamental Sciences (IPM), School of Computer Science, Tehran, 19395-5531, Iran}
\affil[3]{University of Waterloo, Cheriton School of Computer Science, Waterloo, N2L3G1, Canada}
\affil[*]{mina.zarei@iasbs.ac.ir}

\begin{abstract}
The collective behaviour of neural networks depends on the cellular and synaptic properties of the neurons. The phase-response curve (PRC) is an experimentally obtainable measure of cellular properties that quantifies the shift in the next spike time of a neuron as a function of the phase at which stimulus is delivered to that neuron. The neuronal PRCs can be classified as having either purely positive values (type I) or distinct positive and negative regions (type II). Networks of type 1 PRCs tend not to synchronize via mutual excitatory synaptic connections. We study the synchronization properties of identical type I and type II neurons, assuming unidirectional synapses. Performing the linear stability analysis and the numerical simulation of the extended Kuramoto model, we show that feedforward loop motifs favour synchronization of type I excitatory and inhibitory neurons, while feedback loop motifs destroy their synchronization tendency. Moreover, large directed networks, either without feedback motifs or with many of them, have been constructed from the same undirected backbones, and a high synchronization level is observed for directed acyclic graphs with type I neurons. It has been shown that, the synchronizability of type I neurons depends on both the directionality of the network connectivity and the topology of its undirected backbone. The abundance of feedforward motifs enhances the synchronizability of the directed acyclic graphs.
\end{abstract}
\begin{document}

\maketitle
%
%
\thispagestyle{empty}
\section*{Introduction} 

For several decades, there has been a continuing research interest in the synchronization phenomenon due to its widespread application in natural and artificial systems ranging from neural dynamics~\cite{belykh2005synchronization,gregoriou2009high,wang2011synchronous}, cardiac pacemaker cells~\cite{glass1988clocks}, and power grid networks~\cite{motter2013spontaneous} to social networks~\cite{chowdhury2019synchronization}. Particularly, synchronization plays a key role in the proper functionality of brain neurons. The synchronization of neuronal networks has been associated with many cognitive processes including  memory formation~\cite{axmacher2006memory}, directed attention~\cite{missonnier2006frontal, tiitinen1993selective}, and the processing of sensory stimuli~\cite{bartlett1959synchronization}. 
	
Synchronization patterns mainly depend on the dynamical properties of individual oscillators as well as the underlying structural connectivity. It has been found that brain neurons have different types of intrinsic dynamics that close to threshold, they may be grouped into two excitability classes: Type I and Type II. These two types of neurons are  different in terms of the bifurcation structure observed during their transition to firing. Type I oscillations arise via a saddle-node-onto-limit-cycle bifurcation, whereas type II oscillations are  initiated by Andronov-Hopf bifurcation. In addition, type I neurons exhibit a continuous frequency-current curve, whereas type II neurons show a discontinuous frequency-current curve~\cite{ermentrout1996type, izhikevich2007dynamical}.

The aforementioned excitability type of a single neuron can be quantified by the phase response curve (PRC) which is an experimentally obtainable measure based on the transient change in the cycle period of the neuron in response to an external stimulus~\cite{schultheiss2011phase}.  Different profiles of the neuronal phase response curve arise for different types of neurons  in that in type I neurons, any excitatory perturbation causes an acceleration of the next spike, while in type II neurons, perturbations cause acceleration or delay of the next spike depending on the phase at which the perturbation is delivered to that neuron. These qualitatively different responses to stimulation lead to dramatically different synchronization patterns in  neural networks. Previous studies have shown that, type I cells exhibit relatively poor propensity for synchronization under excitatory couplings,  while type II cells are synchronized better~\cite{ermentrout2001effects,smeal2010phase,ermentrout1996type,bolhasani2015stabilizing}. Experimental results verify the neurons' ability to switch between types as well as the coexistence of both cell types in the brain~\cite{reid2004firing,stiefel2009effects,stiefel2008cholinergic}.  For instance, changes in the neocortical synchronization during sleep-wake cycle are often associated with excitability changes in the cortical neurons~\cite{bear_corners_paradiso_2016,platt2011cholinergic}.

Regarding the underlying network connectivity, Master Stability Function is a well-known formalism providing estimation of the synchronization stability in the networks of coupled identical oscillators~\cite{pecora1998master}. According to this formalism, the spread of the eigenvalues of the Laplacian matrix is a synchronizability index. That is, the more compact the eigenvalues, the more likely the synchronous state will be stable. This index works well when comparing synchronizability of chaotic oscillators and identical Kuramoto oscillators. Using the index, it can be shown that complete graphs and  directed acyclic graphs with identical node in-degrees are optimal networks provided that they embed an oriented spanning tree~\cite{nishikawa2006synchronization,nishikawa2006maximum}. 

An entirely different approach to explore the impact of network structure on synchronization is investigating important and dominant topological features of networks known as motifs~\cite{d2008synchronization,lodato2007synchronization,soriano2012synchronization,esfahani2016stimulus}. Motifs are significantly over represented subgraphs that have been recognized as the building blocks of many real-world networks in various domains~\cite{sporns2004motifs}. Therefore, it’s important to understand how the global dynamics are affected by the network motifs. For example, previous studies have shown that feedback loop motifs contribute to  two phenomena: responses to noise and dynamical stability~\cite{sheshbolouki2015feedback,gerard2012effect}.

In this paper, we investigate the role of network motifs in synchronization of directed neural networks with type I or type II neurons. Among various network motifs, we  focus on two distinct categories: (I) feedback loops (FBL) and (II) feedforward loop motifs (FFL). These motifs are commonly highlighted features observed in real-world networks and have attracted a great deal of attention in the literature~\cite{mangan2003structure,alon2007network}. In order to study the dynamics of a large collection of such motifs, we generate two different directed graphs, with the same undirected skeleton, constructed only from FFLs or FBLs. These directed graphs are called directed acyclic graph (DAG) and balanced directed graph (BDG), respectively. Both analytical and simulation results show that type I  excitatory and inhibitory neurons are synchronized in directed acyclic networks (i.e. networks without any feedback loops). However, these neurons fail to synchronize in undirected networks. In fact, feedback loops destroy synchronization of the networks with identical type I neurons. 

\section*{Material and Methods}
\subsection*{Structure of the networks}

An orientation of an undirected graph is an assignment of a single direction to each of its edges, turning the initial graph into a directed graph. In this paper, using two different algorithms, we built different orientations of a graph such that the number of feedback loops (FBLs) is  maximized in one orientation and minimized in the other one. These algorithms are described bellow.

It has been shown that, the number of feedback loops increases by enhancing the correlation between the in- and out-degrees of the nodes~\cite{bianconi2008local}. Therefore, we start with an undirected graph, and try to assign each edge a single direction in such a way that the in-degree and the out-degree of any given node are almost the same. The Eulerian cycle in an undirected graph is a cycle that visits every link exactly once. An undirected graph has an Eulerian cycle if and only if every vertex has an even degree. Eulerian orientation of a graph is an orientation that directs the edges along the Eulerian path such that every vertex has the same number of incoming and outgoing edges. The resulting directed graph is called a balanced directed graph (BDG) and has maximum number of feedback loops. In a network in which some nodes are of odd degree, we can still build an orientation in a way that the resulting directed graph is almost balanced.  To this end, a new fake node is added to the undirected graph and any existing node with an odd degree is connected to this new node. Since every undirected graph has an even number of nodes with odd degree, all the nodes in the new undirected graph have even degree and we can build the Eulerian orientation of the new graph. After orienting the graph, the fake node is removed with all its adjacent links. In this manner,  the difference between in- and out-degree of each node is at most one and the oriented graph is almost balanced~\cite{zarei2009complex}.

Directed acyclic graphs (DAGs) are constructed by extending the residual degree gradient  method~\cite{son2009dynamics,son2009dynamics,sheshbolouki2015role}. At first, each node is labeled by a residual degree which determines the number of adjacent undirected links. The residual values drop as more adjacent links are assigned directions. Thus, initial and final residual values for each node equal to its degree and zero, respectively. In each iteration of assigning the directions, the node with the smallest residual degree is selected and all of its adjacent undirected links are assigned incoming directions. The residual values are updated afterwards and iterations are repeated until there are no undirected edges left in the network.  Since  each node will have incoming links only from the previously chosen nodes, the constructed graph will be acyclic. It is probable that this method gives rise to a disconnected DAG which in turn leads to an incomplete synchronization. Therefore, to avoid drawing incorrect conclusions when comparing synchronizability of different graphs, we have considered just connected DAGs in our simulations.

It should be noted that, directed and undirected networks should have the same number of nodes and arcs (i.e. unidirectional links) to be comparable in terms of their synchronizability.  To this end, we double the number of arcs while giving directions to the links of an undirected graph. Therefore, the adjacency matrix of an undirected graph is defined as  $a_{ij} = 1$, if nodes $i$ and $j$ are connected, and $a_{ij}$ = 0, otherwise. Whereas, the adjacency matrix of a directed graph is defined as, $a_{ij}=2$, if a link is directed from node $j$ to node $i$, and $ a_{ij} = 0$, otherwise.
\subsection*{Phase response curve (PRC)}
\begin{figure*}[t]
	\centering
	\includegraphics[scale=0.3, trim=0 30 0 20]{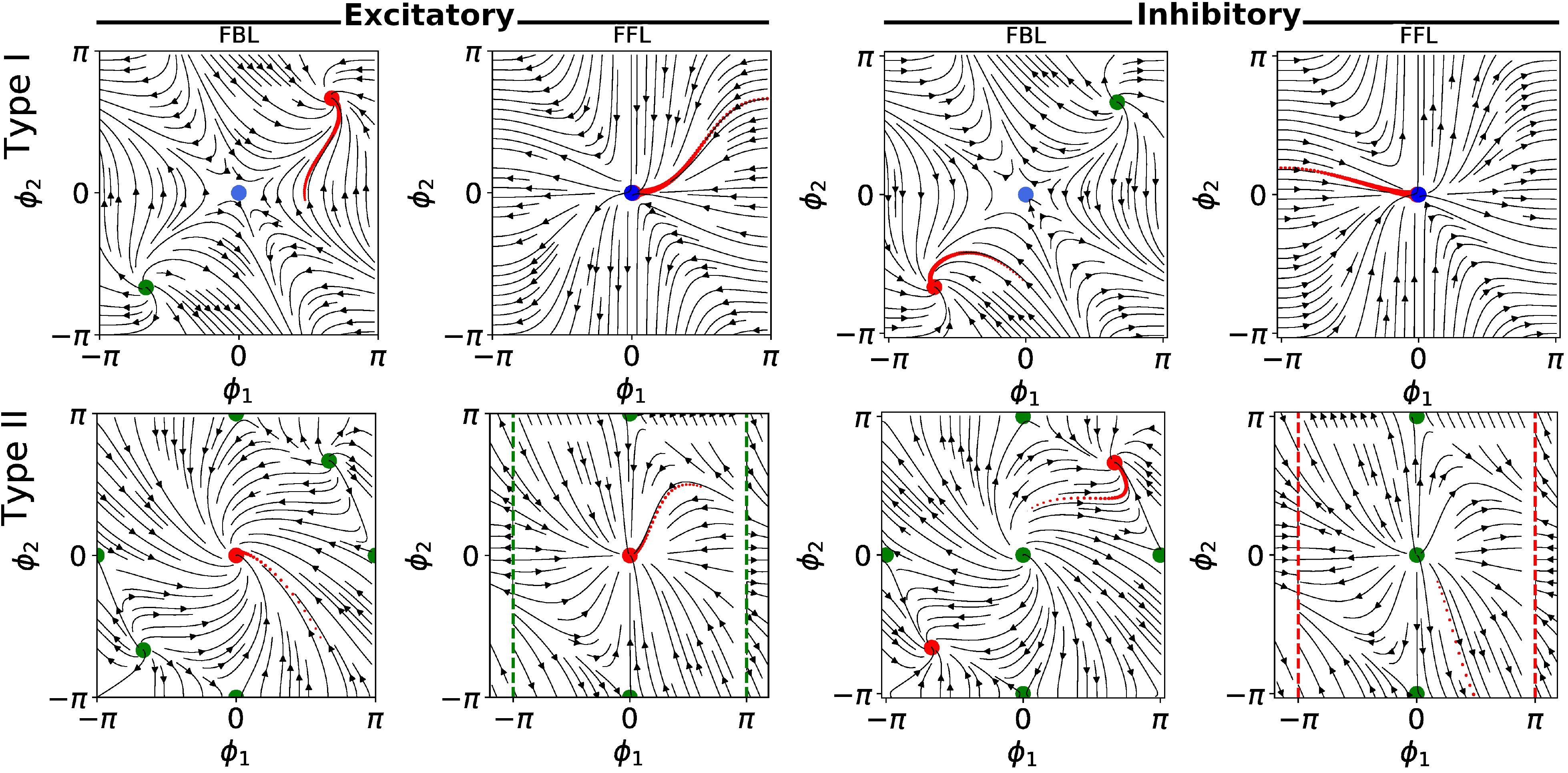}
	\caption{\textbf{ Phase plane portraits for reduced two-dimensional systems of type I and type II motifs.} The phase plane portraits for reduced two-dimentional systems of (top row)~type I and (bottom row)~type II  phase oscillators situated in a FBL, and FFL motifs.  The plots on the left-hand side (right-hand side)  of the figure correspond to the excitatory (inhibitory) oscillators. The stable and unstable (or saddle) fixed points have been distinguished by red and green colours, respectively. The Jacobian matrices of the fixed points coloured in blue have two zero eigenvalues. }
	\label{plane}
\end{figure*}
\begin{figure}[t]
    \centering
	\includegraphics[scale=0.3, trim=0 30 0 0]{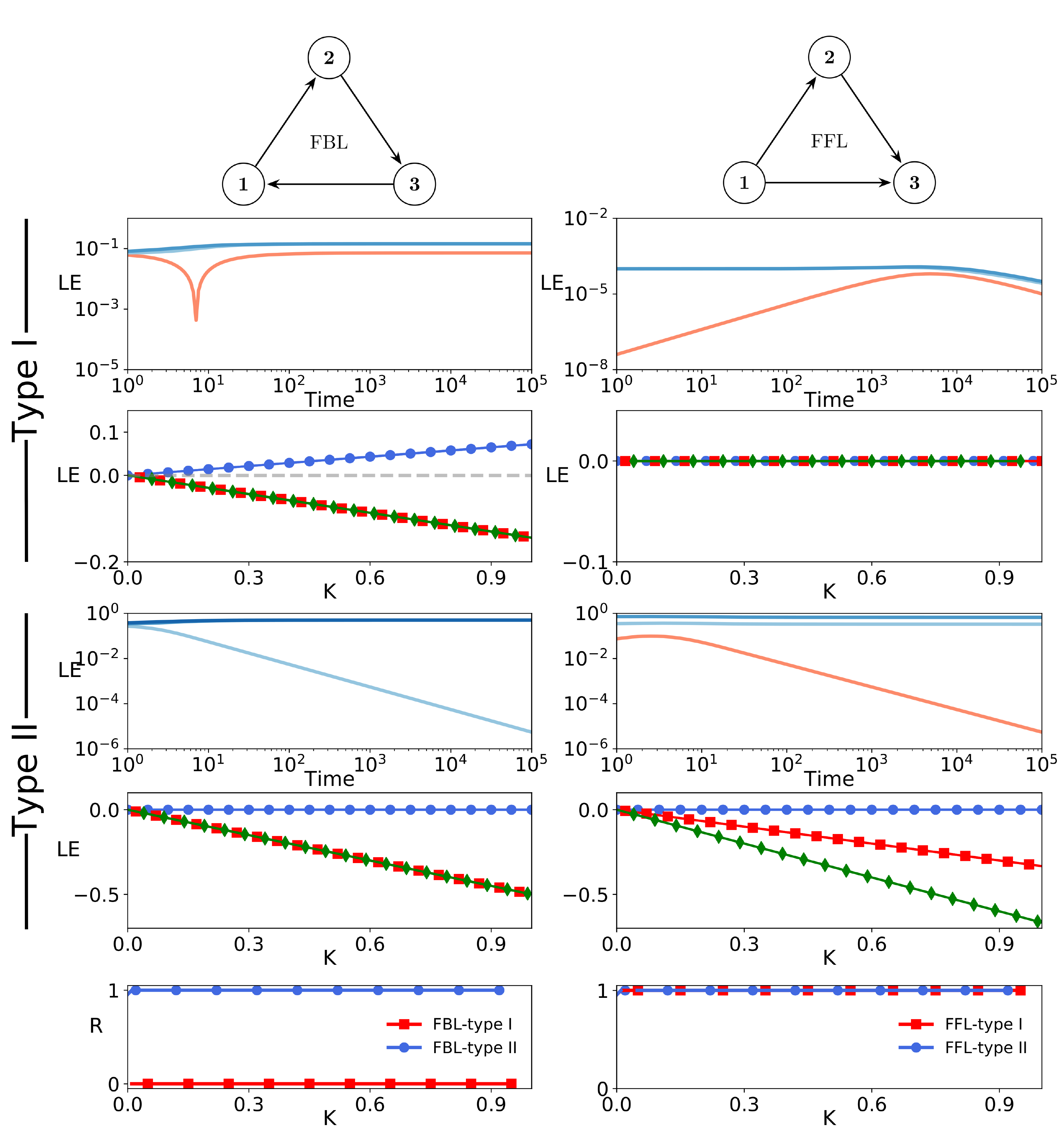}
	\caption{\textbf{Synchronization stability of excitatory FBL and FFL motifs.} (Top four rows)~ the Lyapunov exponents versus time (log-log scale) and the coupling for identical type I and type II excitatory phase oscillators connected by FBL~(left column) and FFL loops~(right column). The sign of Lyapunov exponents in the log-log plots are colour coded, where blue and red indicate negative and positive values, respectively.  (Bottom row)~The stationary order parameter versus coupling for identical type I and type II phase oscillators connected by FBL~(left column) and FFL loops~(right column).}
	\label{loops}
\end{figure}

\begin{figure}[t]
    \centering
	\includegraphics[scale=0.3, trim=0 30 0 0]{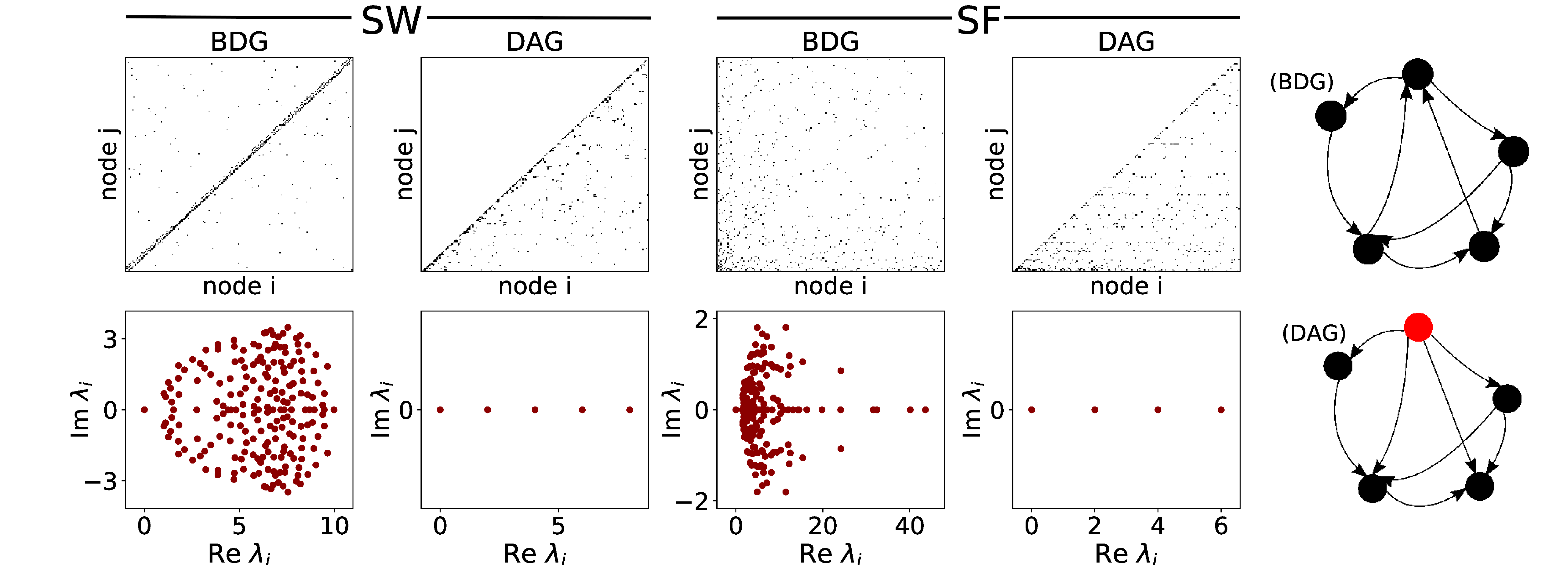}
	\caption{\textbf{Structural properties of two different directed networks.} Adjacency matrices~(top)  and the location of Laplacian eigenvalues in the complex plane~(bottom), for small-world  and scale-free BDGs and DAGs with 200 nodes and average degree of 6. Schematic illustrations of  BDGs and DAGs constructed from the same undirected backbone are shown at the right.  The source node is highlighted with red colour.}
	\label{directioning}
\end{figure}
\begin{figure}[t]
    \centering
	\includegraphics[scale=0.29, trim=0 30 0 0]{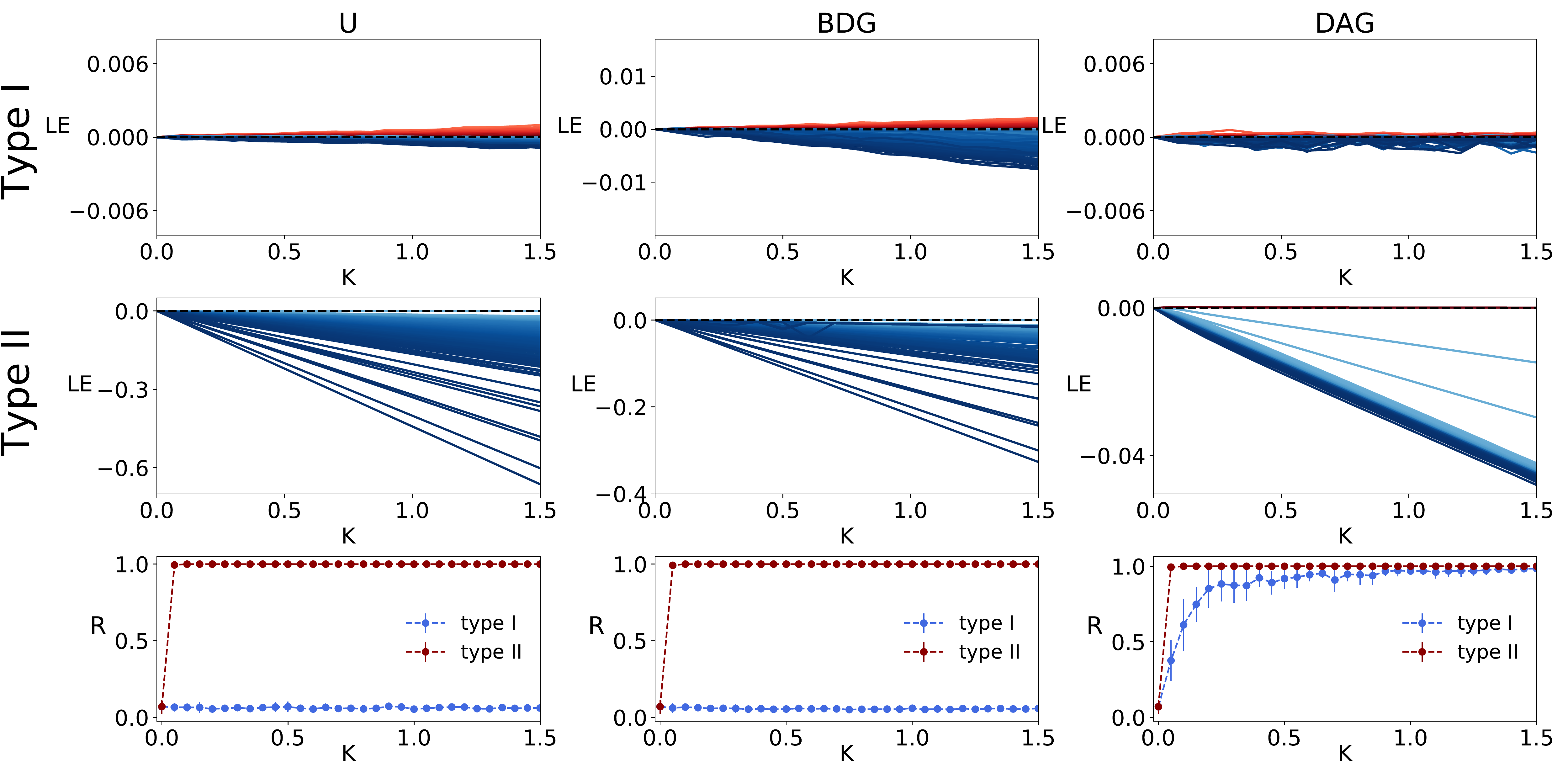}
	\caption{\textbf{ Differences in the synchronization stability of excitatory scale-free networks with various directionalities and neuronal types. }  Lyapunov exponents versus coupling for (top row)~identical type I  and (middle row)~type II  phase oscillators situated on (left column)~scale-free undirected graphs, (middle column)~BDGs and (right column)~DAGs, with \textbf{N=200} nodes are illustrated. The sign of the Lyapunov exponents is colour coded, where blue and red indicate negative and positive values, respectively. (Bottom row)~The stationary order parameter versus the coupling strength of the  scale-free networks with different link-directionalities (i.e. U(Undirected), BDG and DAG) are illustrated.}
	\label{bdg-dag}
\end{figure}
\begin{figure}[t]
	\centering
	\includegraphics[scale=0.35, trim=0 30 0 0]{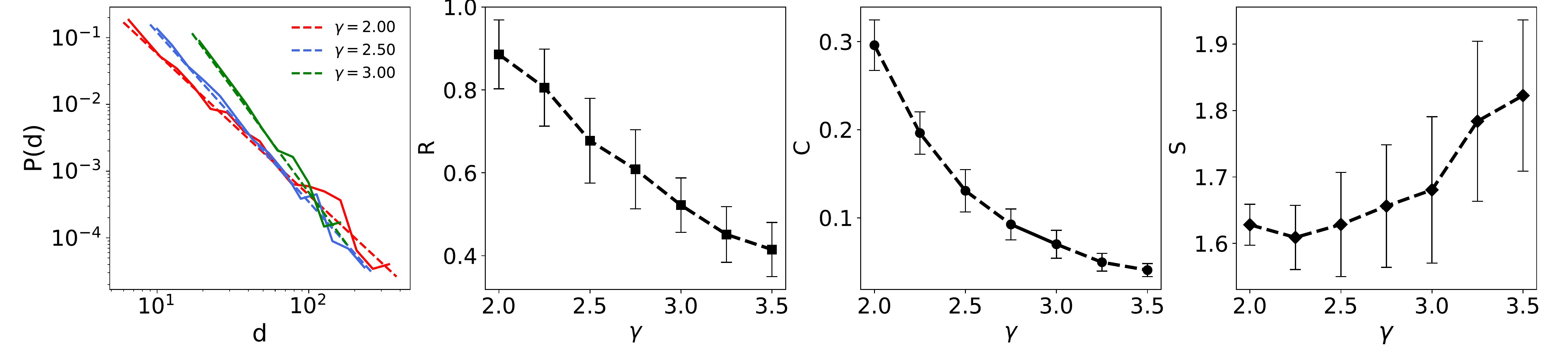}
	\caption{\textbf{Comparison of synchronization of type I excitatory phase oscillators on excitatory scale-free DAGs with different exponents. } (First Column)~The degree distribution of the scale-free networks with different scaling exponents, 1024 nodes and average degree of 10. Dashed lines represent linear curve fittings to the probability distribution functions. (Second Column)~The  stationary order parameters of scale-free DAGs versus scaling exponents, for $K=3$. (Third and fourth Columns)~display the average clustering coefficient ($C$) of the undirected backbones and the average shortest path from the source node to other nodes ($S$), versus the scaling exponents. The results are averaged over 50 realizations.}
\label{gamma}
\end{figure}

The phase response curve (PRC) is an illustrative tool to determine the phase shift of an oscillating neuron in response to a brief current pulse delivered at various phases of the cell cycle~\cite{smeal2010phase}.  It can be defined as, $PRC = 1-\frac{T_{\theta}}{T_0}$. Where $T_0$ is the unperturbed cycle period of the neuron, and $T_{\theta}$ is the cycle period when the neuron is perturbed at phase $\theta$. Therefore, positive and negative values of PRC indicate phase advances and delays, respectively.
As  mentioned before,   neurons are classified into two excitability classes: type I and type II. Type I neurons always respond to a small  excitatory stimulus by advancing the next spike. Therefore, they have positive values of PRC for all phases.  In contrast, type II neuron, can advance or delay the next spike depending on the phase at which the perturbation is delivered to them.  Therefore, for type II neurons,  the PRC versus $\theta$ diagram,  contains both positive and negative regions. Studies verify that the Hodgkin-Huxley neurons have PRC II excitability type, while the Wang-Buzs{\'a}ki and Traub neurons have PRC I excitability type~\cite{mato2008type}.
\subsection*{Dynamics of the neurons}
Neurons can be modeled using both phase oscillators and conductance-based models. Since phase models are simple enough to be mathematically tractable, the networks of coupled phase oscillators have been widely studied to model biological and physical systems including neuronal networks. On the other hand, the detailed investigation of the neuronal dynamics and systematic parameter variation is possible through the use of conductance-based models. Here, we choose the extended Kuramoto model (simulating dynamics of coupled phase oscillators) and  Wang-Buzs{\'a}ki and Traub neuron model (simulating the dynamics of coupled spiking neurons) to consider different aspects in the synchronization of neuronal networks. In the following, each model is explained with its corresponding synchronization order parameter.
\subsubsection*{Phase model}
We use the extended Kuramoto model~\cite{kuramoto2003chemical,mofakham2016interplay} to analyse the dynamics of directed networks with type I or type II oscillators. This model assumes a network as a collection of $N$ coupled phase oscillators such that the evolution of the $i$th oscillator is given by:
\begin{equation}
 \dot{\theta}_i=\omega_i+\frac{K}{N} \displaystyle {\sum_{j=1}^{N} a_{ij} G(\theta_i,\theta_j)} ,\hspace{0.3cm}G(\theta_i,\theta_j)=u_i\sin(\theta_j-\theta_i)+(1-u_i)\frac{(1-cos(\theta_j-\theta_i))}{2}\\
\label{kuramoto}
\end{equation}
where $\theta_i$ is the phase of the $i$th oscillator and $K$ is the overall coupling strength. The adjacency matrix of the graph is given by $A = (a_{ij})$. Type I and type II oscillators are distinguished by $u_i=0$ and $u_i=1$, respectively. The function $G(\theta)$ corresponds to the phase response curve of the oscillator (see the \textbf{figure~S1)}. In our simulations, the initial values of $\theta_i$ are randomly drawn from a uniform distribution in the interval $[0,2\pi]$, and natural frequencies are identical. The degree of synchrony of the phase oscillators is quantified by the Kuramoto order parameters $r$,  which is defined as $r(t)= {\frac{1}{N}} {\langle | \sum_{i=1}^{N} \mathrm{e}^{i\theta_{i}(t)}| \rangle}$. 
Here, $\langle\dots\rangle$ represents averaging over different network realizations and initial conditions. The magnitude is $0 	 \leq r \leq   1$. The extreme cases are $r=1$  (coherent state) and $r=0$ (incoherent state).  The time average of $r$ after achieving a steady state is symbolized by $R$. 

\subsubsection*{Neuronal model}
We used Wang-Buzs{\'a}ki model to analyse the dynamics of the networks with type I inhibitory interneurons. The Wang-Buzs{\'a}ki neuron model includes the following differential equation~\cite{wang1996gamma}:
\begin{equation*}
\begin{aligned}
& C_m \frac{dv}{dt}  = -I_{Na}-I_K-I_L-I_{syn}+I_{app}, \hspace{0.25cm} \frac{dh}{dt} = \phi (\alpha_h (1-h)-\beta_h h ),\hspace{0.25cm}\frac{dn}{dt} = \phi (\alpha_n (1-n)-\beta_n n ),\hspace{0.25cm} \frac{ds}{dt} = \alpha F(v)(1-s)-\beta s. 
\end{aligned}
\end{equation*}
where $C_m $ is the capacitance of the neuron and $v$ is the neuron's membrane potential.   $h$ and $n$ are time-varying activation variables that depend on voltage-dependent rate functions ( $\alpha_h$, $\beta_h$,
$\alpha_n$, $\beta_n$). $s$, $\alpha$ and $\beta$ are activation variable and rate functions for synapses.  $\phi$ is the control parameter which directly alters the time constants of sodium inactivation and potassium activation. $F$ is a scaling factor which is given by $F(v)=1/(1+\exp(-(v-\theta_{syn})/2))$. Here, $\theta_{syn}$ is a parameter which must be set high enough to have neurotransmitter release only when a spike has been emitted in the presynaptic neuron.   $I_{Na}$, $I_K$, $I_L$, $I_{syn}$, and $I_{app}$  represent sodium ($Na^+$), delayed rectifier potassium ($K^+$), leakage,  synaptic and external currents, respectively. The currents are modeled by the following equations:
\begin{equation*}
 I_L    = g_L(v-E_L),\hspace{0.2cm}I_{Na} = g_{Na}m_{\infty}^3h(v-E_{Na}),\hspace{0.2cm}I_K    = g_K n^4 (v-E_K),\hspace{0.2cm}I_{syn,i} = \frac{K}{N}\sum_{j=0}^{N_i} g_{syn}s(v_j-E_{syn}).
\end{equation*}
 where $g_{L}$, $g_{Na}$, $g_K$, and $g_{syn}$ are the maximal values of the conductances for leak, sodium, potassium and synaptic currents, respectively, and  $E_L$, $E_{Na}$, $E_K$, and $E_{syn}$ are their respective
reversal potentials. $K$ is the coupling strength and the summation in $I_{syn,i}$ occurs over all neighbours of the $i$th neuron. The steady state  activation parameter of the sodium current is $m_\infty = \alpha_m/(\alpha_m+\beta_m)$. The parameters of the Wang-Buzs{\'a}ki model~\cite{wang1996gamma} are summarized in the table~S1. The rate functions are given by the following functions: 
\begin{equation*}
\begin{aligned}
& \alpha_m= -0.1(v+35)/(\exp(-0.1(v+35))-1),\hspace{0.1cm}\alpha_h= 0.07 \exp(-(v+58)/20),\hspace{0.1cm}\alpha_n= -0.01(v+34)/(\exp(-0.1(v+34))-1),\\
& \beta_m = 4 \exp(-(v+60)/18), \hspace{0.1cm} \beta_h = 1/(\exp(-0.1(v+28))+1), \hspace{0.1cm} \beta_n= 0.125 \exp(-(v+44)/80).
\end{aligned}
\end{equation*}

We have also used  the the Traub pyramidal excitatory type I neuron model~\cite{traub1982simulation}. The equations describing the model is $ C_m \frac{dv}{dt}  = -I_{Na}-I_K-I_L-I_{syn}+I_{app}$, where gating variables $m,n$ and $h$ obey equations of the type $\dot{x}=\alpha_x(v)(1-x)-\beta_x(v)x$. The rate functions are:
\begin{equation*}
\begin{aligned}
&\alpha_m= 0.32 (v+54)/(1-\exp(-(v+25)/4)), \hspace{0.1cm} \alpha_h= 0.128 exp(-(v+50)/18), \hspace{0.1cm} \alpha_n = 0.032(v+52)/(1-\exp(-(v+52)/5))\\
&\beta_m= 0.28 (v+27)/(1-\exp((v+27)/5)-1), \beta_h= 4/(1+\exp(-(v+27)/5)), \hspace{0.1cm}\beta_n = 0.5 \exp(-(v+57)/40).
\end{aligned}
\end{equation*}
Synaptic gates satisfy $\dot{s}=\alpha(1-s)/(1+\exp(-(v+10)/10))-\beta s$, and synaptic current to the \textit{i}th neuron is evaluated from $I_{syn,i}=K\sum_{j\in N_i} g_{syn}s(V_{res})$, where $N_i$ are neighbors of the \textit{i}th neuron and $V_{res}$ is resetting membrane potential.  The parameters of the  Traub neuron model, which corresponds to AMPA-receptor-mediated synapse, are summarized in the table~S2. 

There are various measures to quantify the level of synchrony in a large population of neurons within a network. Among the various metrics, we choose the interspike distance synchrony measure, denoted as $B$. As one would expect, the minimum interval between spikes of different neurons in  synchronized state is less than that of asynchronous state. Based on this idea, the spike synchrony measure is defined as follows~\cite{tiesinga2004rapid}: 
\begin{equation}
B = \bigg( \frac{\sqrt{\langle \tau_x^2 \rangle_{\tau} - \langle \tau_x \rangle_{\tau}^2 }}{\langle \tau_x \rangle} -1 \bigg) \frac{1}{\sqrt{N}}
\end{equation}
where,  $N$ is the number of neurons in the network and $\tau_x = t_{x+1} - t_x$ denotes the interspike interval of the merged set of network spikes.  Here, the inter-spike intervals are calculated between different neurons. $\langle\dots\rangle_{\tau}$ denotes the averaging over all intervals.   It can be shown that $B$ is bounded between 0 and 1, where  $B=1$ and $B=0$ represent  fully synchronized and asynchronous states, respectively. The voltage synchrony~\cite{lim2009coupling} is another commonly used measure that has been investigated in the supplementary material. 
\section*{Results}
First, we studied the synchronization stability of the FBLs and  FFLs connecting identical type I or type II phase oscillators analytically. For different motifs, we considered the reduced two-dimensional systems of the  phase differences (the equations are provided in the supplementary material). As expected, we found stable (unstable) synchronous states for type II excitatory (inhibitory) FFLs and FBLs. On the other hand, we observed stable asynchronous state for type I excitatory and inhibitory FBLs, and stable synchronous state for type I excitatory and inhibitory FFLs. In fact, both eigenvalues of the Jacobian matrix, for the synchronous state of the type I FFL, are zero. Consequently,  this fixed point is globally stable, but not asymptotically. It means that, if one perturbs the system and then leaves the system alone, the time dependence of the perturbation is milder than exponential. The phase portraits of the type I and type II excitatory and inhibitory oscillators situated in different motifs are plotted in the \textbf{figure~\ref{plane}}. 

We also numerically investigated the synchronization stability of the type I and II identical excitatory phase oscillators including different motifs (detailed explanation of the method is found in the supplementary material and the article written by Wolf~{\it et al}~\cite{wolf1985determining}). The evolutions of the three Lyapunov exponents for the type I and the type II FBLs and FFLs are demonstrated in the first and the third rows of the \textbf{figure~\ref{loops}}. The steady Lyapunov exponents versus coupling strengths are plotted at the second and the fourth rows of the \textbf{figure~\ref{loops}}. We can see that, as we expected from the analytical results,  all the three Lyapunov exponents for the type I FFLs are zero.  The bottom panels of the \textbf{figure~\ref{loops}} illustrate the evolution of the stationary order parameter for the type I and II excitatory phase oscillators connected by FBLs and FFLs with various coupling strengths. As anticipated, the results display the stable synchronous state for the type II FBLs and FFLs, and the type I FFLs. Therefore, the feedforward loop motif is an optimal structure for synchronization of the identical excitatory type I phase oscillators. This is also correct  for the identical inhibitory type I phase oscillators (see \textbf{figure~\ref{plane}}).

\textbf{Figure~\ref{directioning}} shows the structural properties of the BDGs and DAGs constructed from the same undirected Watts-Strogatz small-world and Barab{\'a}si-Albert scale-free networks~\cite{watts1998collective,barabasi1999emergence}. The adjacency matrices have been exhibited in the first row. The adjacency matrix of a directed acyclic graph with a single source node (i.e. a node with zero in-degree) can be transformed into strictly lower triangular form by a permutation matrix.  Therefore, as we can see,  our orientation algorithm constructs DAGs correctly from different undirected graphs. The Laplacian matrix of directed network is defined as $L=D_{in}-A$, where $D_{in}$ is a diagonal matrix of in-degrees and $A$ is the adjacency matrix.  Real and imaginary parts of the eigenvalues of the Laplacian matrices are shown in the bottom panels. The real parts  and the imaginary parts of the complex eigenvalues have their own information as the real parts reflect the undirected topology of the network and the imaginary parts reflect the effects of directed links and loops~\cite{caughman2006kernels,zarei2009complex}. For example, since the eigenvalues of a lower triangular matrix are the same as its diagonal entries, the Laplacian eigenvalues of a DAG are real and  equal to the in-degrees of its nodes. Therefore, as we expected DAGs do not have any FBL and their eigenvalues are real. On the other hand, we can see that the variance of the imaginary part of the  small-world BDGs is higher than that of the scale-free BDGs. It is because of the fact that clustering coefficient of a small-world network is higher than clustering coefficient of scale-free networks with the same number of nodes and arcs. Therefore, small-world BDGs have higher number of FBLs than scale-free BDGs. In addition, because of the hubs (i.e. large-degree nodes in the undirected graphs), the real parts of the eigenvalues of the scale-free BDGs are more spread out than small-world BDGs. On the other hand, the real parts of the eigenvalues of the scale-free DAGs are compacter than small-world DAGs, because they have narrower in-degree distributions. Note that based on our orientation algorithm the source node would be a hub in a scale-free DAG.  Now that we understood the structural differences between our directed networks, in the following we will compare the synchronizability of type I and type II oscillators on these structures. 

As mentioned,  unlike FBLs, FFLs favour synchronization in networks of type I oscillators. However, in real world networks motifs are merged together to form more complex structures. In order to see the effects of FFLs and FBLs in the synchronization of the large networks, we constructed two different directed networks from the same undirected backbone. Balanced Directed Graphs (BDGs) with many FBLs, and  Directed Acyclic Graphs (DAGs) without any FBL. The flow of information in the BDGs is similar to their undirected backbone, but the information flow in the DAGs is directed from the source node to the rest of the graph. Therefore, BDGs and DAGs are very different, even if they have the same undirected skeleton.

The synchronizability of the scale-free networks with different link-directionalities has  been investigated in the \textbf{figure~\ref{bdg-dag}}.  We can see that type II oscillators are synchronized on all the three network structures. However, type I excitatory oscillators can be synchronized only on DAGs. This proves that FBLs are harmful for synchronization of type I oscillators, while FFLs enhance their synchronization.  In addition, we found similar synchronization level for BDGs and undirected graphs. This result is in agreement with previous finding that the synchronization condition for the symmetrized network guarantees synchronization in the asymmetrical network with node balance~\cite{belykh2006synchronization}. According to the plot of the Lyapunov exponents versus the coupling strengths, all the Lyapunov exponets of the type I DAG are small. Therefore, we considered very long time simulation to estimate the Lyapunov exponents of type I DAG. The results confirm that the absolute values of  all the exponents are smaller than $10^{-5}$ (see \textbf{figure~S2}). Thus,  the synchronous state for type I scale-free DAGs is globally stable but not asymptotically. This is similar to the synchronous stability of their building blocks, i.e. the FFL motifs (see \textbf{figure~\ref{plane}}). The type I inhibitory phase oscillators are also synchronized well on the DAG structures (see \textbf{figure~S3}).

In order to better investigate the role of undirected topology of the DAGs on their synchronization, we construct the DAGs from scale-free networks ($p(d)\propto d^{-\gamma}$) with different scaling exponents ($\gamma$). As we know, increasing the scaling exponents leads to more degree-homogeneous networks. \textbf{Figure~\ref{gamma}} shows the synchronization of the DAGs constructed from the mentioned scale-free networks. The result shows that increasing the scaling exponents, decreases the clustering coefficient and synchronizabilty of scale-free DAGs. The underlying reason is that, by reducing the scaling exponent in networks with power-law degree distributions, the source node (the node with zero in-degree where the information flow starts from ) would be able to reach every other node in the network easier. This characteristic has been previously shown for synchronization of directed networks of type II oscillators~\cite{sheshbolouki2015feedback}. 

The effect of gradually adding feedback loops to the DAGs is also investigated. The results show that the synchronizability of the directed network decreases nonlinearly with increasing the number of the feedback loops (see  \textbf{figure~S4}).

In reality inhibitory and excitatory neurons work together to perform complex tasks.  It has been suggested that excitatory and inhibitory inputs of a neuron are balanced, and this balance is important for the highly irregular firing observed in the cortex~\cite{haider2006neocortical}. Our results show that disconnected hybrid networks that are formed by feedforward motifs that their effective nodes (i.e. nodes which have outgoing edges) are purely inhibitory or excitatory, can provide balance networks with incoherent dynamics (see \textbf{figure~S5}).
  
\begin{figure}[t]
	\centering
	\includegraphics[scale=0.35,trim=130 10 10 10]{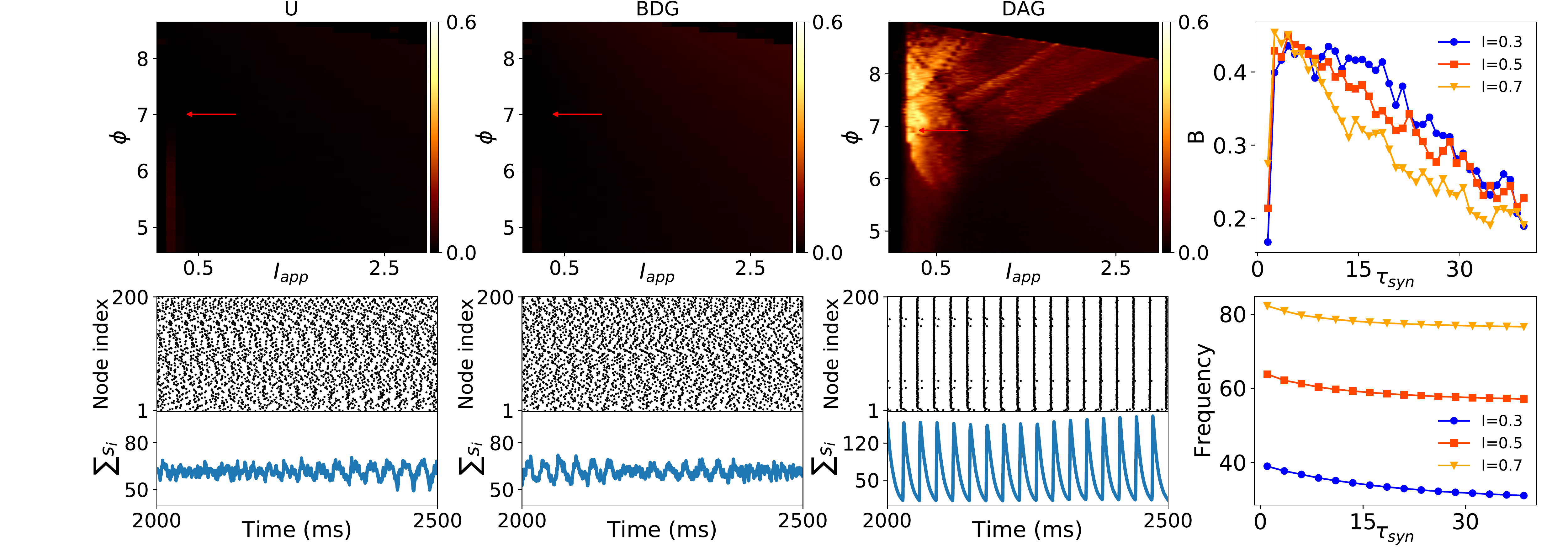}
	\caption{\textbf{Synchronization of Wang-Buzs{\'a}ki inhibitory neurons coupled via scale-free undirected and directed graphs.}  (First three columns)~Top panels display the interspike distance synchrony of the networks on $I_{app}-\phi$ phase space. Bottom panels represent the raster plots and the sum of the synaptic gates $\sum_i s_i$  for three network instances using the parameters specified by the arrows at the top plots. The synapse parameters are $\alpha=12  ms^{-1}, \beta=0.1 ms^{-1}, g_{syn}=0.1 mS/cm^2$. (Right column)~Top and bottom panels show the effects of synaptic time constant ($\tau_{syn} = 1/\beta$) on synchrony and frequency of oscillation for different values of applied currents, respectively at $K=1.25$. The parameters of the networks are valued as $N=200$, and $\gamma=3$.}
	\label{neuron1}
\end{figure}

In the preceding parts, we investigated the synchronization of type I and type II phase oscillators, and we showed that type I excitatory and inhibitory oscillators which are connected by DAGs are synchronized.  However, previous studies have shown that type I neurons are only synchronized with inhibitory synaptic connections~\cite{wang1996gamma}. To address this concern, we study the synchronizability of our oriented networks using conductance-based neuron models. First, we consider the type I  Wang-Buzs{\'a}ki inhibitory neurons connected by scale-free directed networks. The characteristics of the Wang-Buzs{\'a}ki model neurons have been presented in the supplementary \textbf{figure~S6}.  This model has been used to show that gamma rhythm (20-80 Hz), which is observed during behavioural arousal, can emerge in an undirected random network of interconnected inhibitory neurons~\cite{wang1996gamma}. However this rhythmic behaviour is only observed in the dense networks. Top panels of the \textbf{figure~\ref{neuron1}} represent  interspike distance synchrony level of the type I  Wang-Buzs{\'a}ki neurons connected via scale-free undirected (U), BDG, and DAG by inhibitory  synapses in the $\Phi-I_{app}$ parameter spaces. The results verify that type I  Wang-Buzs{\'a}ki neurons are not synchronized on the sparse undirected graph and BDG. In contrast, there exists a parameter region in which neurons can be synchronized in the scale-free DAGs. Moreover, it has been shown that synaptic time constants potentially play a role in the determination of network frequency, and our results confirm this. In fact larger time constant makes the overall decay slower. Therefore,  increasing $\tau_{syn}$ decreases the oscillation frequency for different values of the applied currents. In addition, we can see that the synchrony index displays a peak at the small values of the $\tau_{syn}$. This would mean that synchrony is high only within a  restricted range of frequencies.  These findings are consistent with the previous results obtained for dense undirected networks~\cite{wang1996gamma}.  Here, the decay time constant of 10 ms that we used is consistent with $GABA_A$ receptor~\cite{borgers2017introduction}. The parameter $\phi$ also  affects  the synchronization of the networks. In fact, decreasing $\phi$ reduces the firing rate by producing a deep afterhyperpolarization (see  \textbf{figure~S7}). In our simulation, a spike is detected when the voltage of membrane reaches a fixed threshold ($-55$ mV) from a lower value.

\begin{figure}[t]
	\centering
	\includegraphics[scale=0.35, trim= 130 10 10 10]{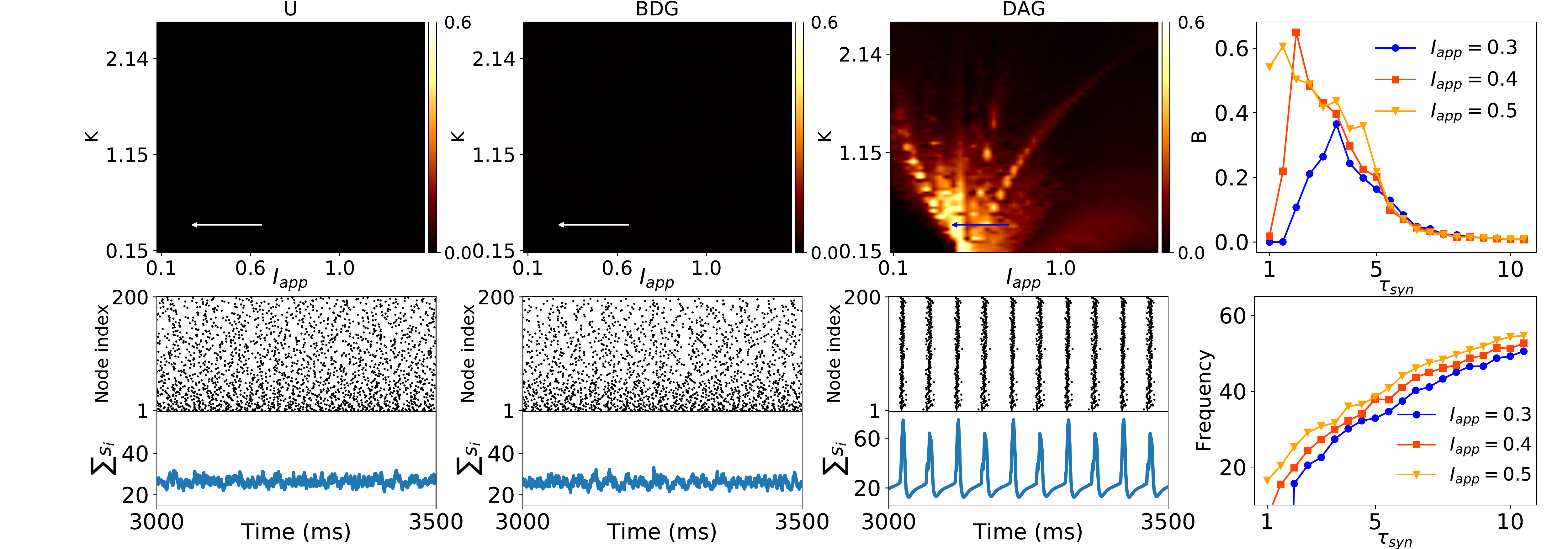}
	\caption{\textbf{Synchronization of Traub neurons coupled via scale-free undirected and oriented graphs.} (First three columns)~Top panels show the spike synchrony of the networks on $I_{app}-K$ phase space. Bottom panels exhibit the raster plots and the sum of the synaptic gates $\sum_i s_i$  for three network instances using the parameters specified by the arrows at the top plots. The synapse parameters are $\alpha=12  ms^{-1}, \beta=0.5 ms^{-1}, g_{syn}=0.01 mS/cm^2$. (Right column)~Top and bottom panels represent the effects of synaptic time constant ($\tau_{syn}=1/\beta$) on spike synchrony and frequency of oscillation, respectively at $K=0.3$. The parameters of the network are valued as $N=200$, and $\gamma=3$.}
	\label{neuron2}
\end{figure}

A similar behaviour is seen for the synchronization of the DAGs constructed by excitatory type I neurons by using Traub neuron model (see  \textbf{figures~\ref{neuron2}} and \textbf{S8} ). However, for excitatory neurons the oscillation frequency increases as the time constant increases.  The results on neuron models are also confirmed by using  voltage synchrony measure (see the \textbf{figures~S9} and \textbf{S10}).

\section*{Discussion}
The observed complex spatiotemporal neural activity patterns reflect both the connectivity and dynamic properties of individual neurons. 
Two types of neuronal activities are found based on their different responses to the small depolarizing current pulses. Type I neurons only speed up the oscillation when they receive the stimulus, whereas type II neurons experience both phase advance and delay, depending on the timing of the perturbation.  Previous studies have shown that, synchonizability of type II neurons is higher than type I neurons in excitatory networks, while inhibition is more stabilizing the synchrony of type I neurons~\cite{whittington2000inhibition,bartos2002fast,chow1998frequency,brunel2006noise,rich2016intrinsic}. Adaptation can stabilize the  synchrony of excitatory type I neurons  by changing their PRC~\cite{ermentrout2001effects}. But without adaptation, excitation is desynchronizing even for fully connected identical type I neurons  beyond the weak coupling regime~\cite{hansel1995synchrony}. However, the master stability formalism suggest that fully connected network is an optimal structure for the synchronization of identical type II neurons. Therefore, the master stability approach can not be applied to investigate the synchronizability of the networks of type I neurons. On the other hand, neurons are connected to each other through synapses which are mostly unidirectional in passing signals. Based on the observations mentioned above, we studied synchronization properties of directed networks of identical type I or type II neurons, assuming excitatory and inhibitory interactions. 

Motifs are significantly over-represented subgraphs of a complex network, and play  important roles in shaping its emergent behaviours. Therefore, firstly we focused on the synchronization properties of two different motifs in directed networks: feedback loops and feedforward loop motifs. Performing the linear stability analysis and the numerical simulation, we showed that synchrony can emerge in the feedforward loops with identical type I oscillators and purely excitatory or inhibitory connections. In fact, this synchronous state is globally stable but not asymptotically. Therefore, a small perturbation to the synchronous state would cause the system to converge to the fixed point milder than exponentially. However, the asynchronous state turns out to be asymptotically stable for feedback loops constructed from identical type I phase oscillators with purely excitatory or inhibitory connections. The analytical results obtained are compared and found to comply with the results obtained by using the extended Kuramoto model simulation. Next, we studied the synchronization of various types of neurons coupled via large directed networks. To this end, we used different methods for assigning the link directions to construct two different directed networks from the same undirected backbone, referred to as balanced degree graph (BDG ) and directed acyclic graph (DAG). Precisely, many feedback loops are merged together to construct a BDG and only feedforward loop motifs are combined to make a DAG. Both the linear stability analysis and the numerical simulation confirmed that identical type I neurons connected by DAGs via strictly excitatory or inhibitory synapses are fully synchronized, while they are not synchronized when they are connected by BDGs. We can see that the steady Lyapunov exponents of type I DAGs are zero. This is in agreement with the result we found for the synchronization of feedforward loops. Therefore, studying  the dynamics of the motifs can provide an intermediate step to better understand the synchronization of the larger collections. Besides the directionality, we showed that the undirected topology of the graphs also affects the synchronization of directed networks. The DAGs constructed from undirected networks with higher clustering coefficients, have higher number of  feedforward motifs and higher synchronizability. In fact,  a DAG is more synchronizable when all of its nodes are easily reachable from the source node. The results are also verified using conductance-based Wang–Buzs{\'a}ki and Traub neuron models. In reality inhibitory and excitatory neurons work together to perform complex tasks. In general, excitatory and inhibitory inputs of a neuron are said to be balanced, and this balance is important for the highly irregular firing observed in the cortex~\cite{haider2006neocortical}. The global dynamic of motifs with hybrid effective nodes is very complex and directly related to the initial phase values. However, Our investigation suggests that excitatory-inhibitory balance and incoherent dynamics can be provided by merging inhibitory and excitatory feedforward motifs.   Investigating the synchronization of neurons with different excitability types is important to understand the function of the diffuse modulatory cholinergic systems in the brain~\cite{bear_corners_paradiso_2016}. Our main result is that edge directionality significantly alter the synchronizability of the type I neurons with purely excitatory connections, and it can not be ignored. Further studies are clearly required to determine the interplay between neuronal excitability, frequency, edge directionality, and network synchronization.

\section*{Acknowledgements}
M.Z wishes to acknowledge financial support of "The Abdus Salam International Centre for Theoretical Physics (ICTP)". 
\section*{Author contributions}
M.Z. conceived the study, A.Z. and A.S. jointly designed and implemented the simulation model, M.Z.
and A.Z. created the analytic results, M.Z. wrote the paper and A.S. edited the manuscript. 

\bibliography{lib}

\end{document}


\title{Supplementary Material}
 
\maketitle

\section*{Phase model} 
\subsection*{ Analytical study of the stability of the synchronized motifs}
In this section, we study the stability of the feedback and feedforward loops constructed from the identical type~I or type~II  phase oscillators. 

We consider the reduced two-dimensional systems of the  phase differences. Therefore, we define $\omega_i=\omega$, $\theta_1-\theta_2 = \phi_1$, $\theta_2-\theta_3=\phi_2$, $\theta_1-\theta_3 = \phi_1+\phi_2$. Two phases $\theta_1(t)$ and $\theta_2(t)$ are synchronized if their difference $\phi_1(t)$ is bounded at the stationary state. 

Using equation~1 (from the main text) and considering {\bf the type~II phase oscillators ($u_i=1$) situated in a FBL}, the phase differences $\phi_1(t)$ and $\phi_2(t)$ clearly satisfy the following equations: 
\begin{equation}
\begin{aligned}
	&\dot{\phi_1} = -\sigma \left[ \sin(\phi_1 + \phi_2) + \sin(\phi_1) \right]~, \\ 
	&\dot{\phi_2} =  \sigma \left[ \sin(\phi_1) - \sin(\phi_2) \right]~,
\label{fbl2}
\end{aligned}
\end{equation}
This system has different fixed points $(\phi^*_1,\phi^*_2)=(0,0),(0,\pm\pi),(\pm\pi,​0), (\pm\frac{2\pi}{3}, \pm\frac{2\pi}{3}) $. Where, the synchronous state ($\phi^*_1,\phi^*_2)=(0,0)$ is stable (unstable) for excitatory (inhibitory having negative $\sigma$) phase oscillators. 

In the case of the {\bf identical type~II  FFLs}, the evolutions of the phase differences are given by the following equations: 

\begin{equation}
\begin{aligned}
	&\dot{\phi_1} = -\sigma \sin(\phi_1)~, \\ 
	&\dot{\phi_2} =  \sigma \left[ \sin(\phi_1) - \sin(\phi_2) - \sin(\phi_1+\phi_2) \right]~,
\label{ffl2}
\end{aligned}
\end{equation}
The system has  one stable (unstable) fixed point ($(\phi^*_1,\phi^*_2)=(0,0)$) for excitatory (inhibitory) oscillators, and two lines of unstable fixed points ($\pm \pi,\phi^*_2$) for both excitatory and inhibitory oscillators. Therefore, according to the above equations identical excitatory type~II motifs are synchronized, but not the inhibitory ones. 

For the {\bf identical  type~I FBLs}, the reduced two-dimensional system is as follows:	
\begin{equation}
\begin{aligned}
	&\dot{\phi_1} = \frac{\sigma}{2} \left[ \cos(\phi_1) -\cos(\phi_1+\phi_2) \right]~, \\
	&\dot{\phi_2} = \frac{\sigma}{2} \left[ \cos(\phi_2) -\cos(\phi_1) \right]~,
\end{aligned}
\end{equation}
This system has three fixed points:  $(\phi^*_1,\phi^*_2)=(0,0), (\pm\frac{2\pi}{3}, \pm\frac{2\pi}{3})$. In the cases of ($(\phi^*_1,\phi^*_2)=(\pm\frac{2\pi}{3}, \pm\frac{2\pi}{3})$), the phases of oscillators  are located at three equidistant points around the unit circle (i.e. asynchronous state). In fact, asynchronous states are stable for excitatory and inhibitory type~I FBLs. 

In the case of the {\bf identical type~I FFLs}, the evolutions of the phase differences are given as follows: 
\begin{equation}
\begin{aligned}
	& \dot{\phi_1} = -\frac{\sigma}{2} (1-\cos(\phi_1))~,\\
	& \dot{\phi_2} = \frac{\sigma}{2} \left[ \cos(\phi_2) -\cos(\phi_1) + \cos(\phi_1+\phi_2) -1 \right]~,
\end{aligned}
\end{equation}
This system has just one fixed point at $(\phi_1^*, \phi_2^*) =(0,0)$. At this fixed point, both eigenvalues of the Jacobian matrix are zero. Therefore,  this fixed point is globally stable, but not asymptotically stable. In fact, if one perturbs the system and then leaves the system alone, the time dependence of the perturbation would be milder than exponential.  Therefore, identical type~I excitatory and inhibitory FFLs are synchronized. The phase plane portraits have been depicted on the \textbf{figure~1} of the manuscript. 
\subsection*{Phase response curve of a phase oscillator}
\begin{figure*}
\centering
	\includegraphics[scale=0.5]{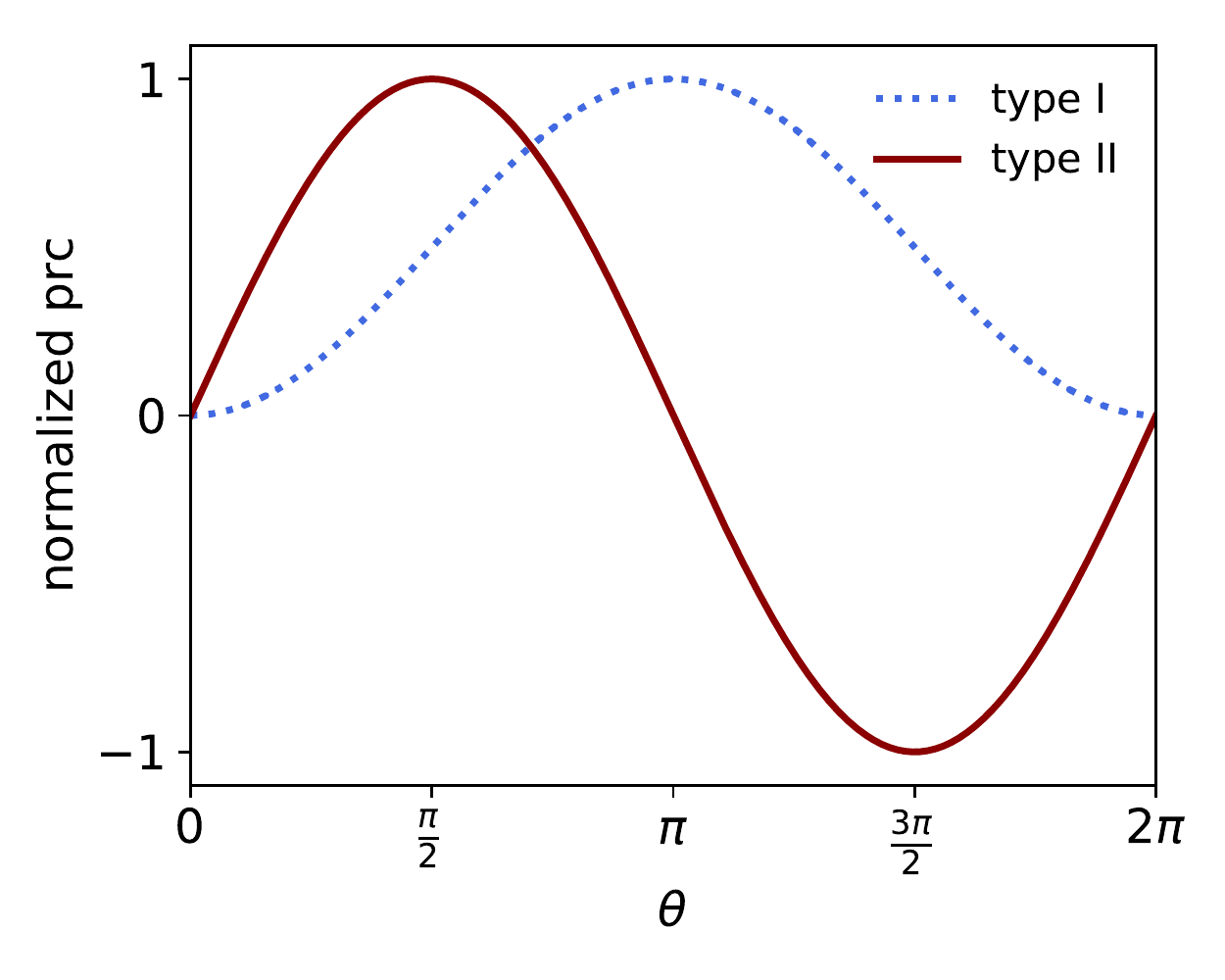}
	\caption{\textbf{Phase response curves of the type~I and type~II phase oscillators.}} 
	\label{prcphase}
\end{figure*}

The phase response curve of a phase oscillator modeled by the generalized Kuramoto model is given by $ G(\theta)=u\sin(\theta)+(1-u)\frac{(1-cos(\theta))}{2}$. Where $u=0$ and $u=1$ correspond to the type~I and type~II oscillators, respectively (see \textbf{figure~\ref{prcphase}}).  
\subsection*{Calculation of the Lyapunov exponents}
Considering our continuous dynamical system in an n-dimensional phase space, we monitor the long-term evolution of an infinitesimal n-ellipsoid of initial conditions. 
The \textit{i}th one-dimensional Lyapunov exponent is then defined in terms of the length of the ellipsoidal principal axis $p_i(t)$~\cite{wolf1985determining}:
\begin{equation}
\lambda_i =  \lim_{t\to\infty} \frac{1}{t} \log_2 \frac{p_i(t)}{p_i(0)},
\end{equation}
where the Lyapunov exponents will be arranged such that: $\lambda_1 \geq \lambda_2 \geq  ... \geq \lambda_n$. The $\lambda_1$ and $\lambda_n$ correspond to the most rapidly expanding and contracting principal axes, respectively.
We simultaneously solve \textit{n} equations of the system and \textit{n $\times$ n} equations of the linearized system. The growth of the corresponding set of vectors is measured, and as the system evolves, the vectors are repeatedly reorthonormalized (e.g. every 50 time steps) using the Gram-Schmidt Orthogonalization (GSO). This makes the vectors maintain a proper phase space orientation.

The simulation programs were written in C++ and using Runge Kutta 4th order integration scheme from Boost odeint library and time step of 0.01 time unit.

The steady-state is determined by convergence test. The simulation program memorizes subsequent values of each lyapunov exponent $\lambda_i$ in \textit{N} buffers of a fixed capacity (e.g 2500 points). When the buffers are full, the program calculates standard deviation of all the values in each of \textit{N} buffers. If the standard deviation of values in any buffer is too high (compared to a fixed threshold e.g. $10^{-6}$), then the buffers are cleared and computations are continued. Otherwise, if the value of standard deviation for each buffer is below a fixed threshold, the calculations are terminated. The final value of $\lambda_i$ returned by the program is equal to the average of all the values $\lambda_i$ memorized in the \textit{i}th buffer~\cite{balcerzak2017fastest}.

\begin{figure*}
	\centering
	\includegraphics[scale=0.24]{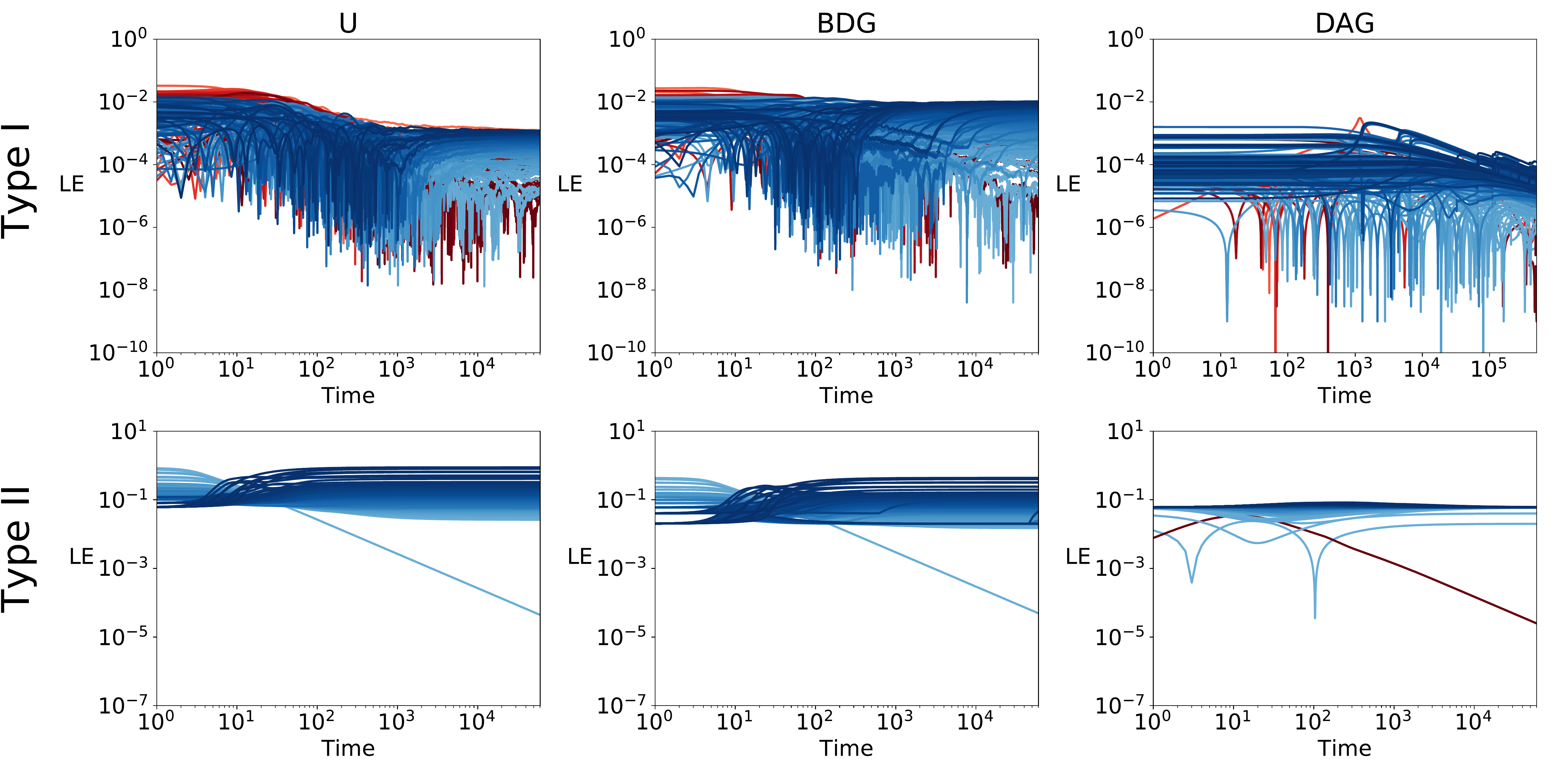}
	\caption{\textbf{The Lyapunov exponents of the extended Kuramoto model on different network structures.} The Lyapunov exponents versus time (log-log scale) for identical (top) type~I and (bottom) type~II excitatory phase oscillators connected by different scale-free networks. The sign of the Lyapunov exponents are colour coded, where blue and red indicate negative and positive values, respectively. }
\label{exponent}
\end{figure*}

The Lyapunov exponents versus time for identical  type~I and type~II excitatory phase oscillators connected by different scale-free networks are depicted in the \textbf{Figure~\ref{exponent}}.

\subsection*{Synchronization of the inhibitory phase oscillators}
\begin{figure*}
	\centering
	\includegraphics[scale=0.3]{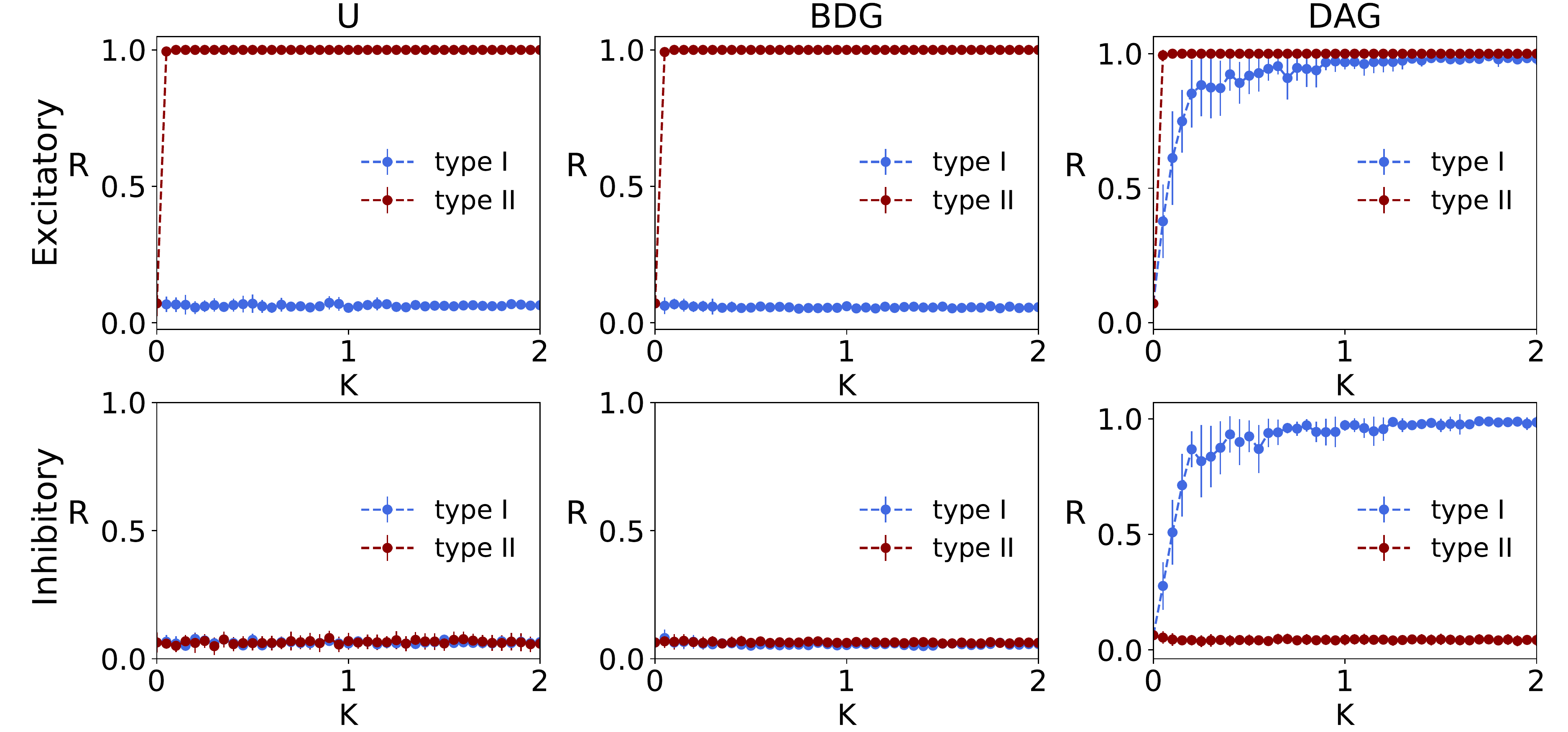}
	\caption{\textbf{Synchronization of the excitatory and inhibitory phase oscillators.} The stationary order parameters versus the coupling strengths for the scale-free undirected and oriented networks with (top)~excitatory and (bottom)~inhibitory phase oscillators.}
\label{inhibitory}
\end{figure*}

\textbf{Figure~\ref{inhibitory}} compares the synchronizability of the excitatory and  the inhibitory scale-free undirected and oriented graphs. We can see that only the DAGs constructed from type~I inhibitory oscillators are synchronized.  

\subsection*{Gradually adding feedback loops to the DAGs}
\begin{figure*}
\centering
	\includegraphics[scale=0.4]{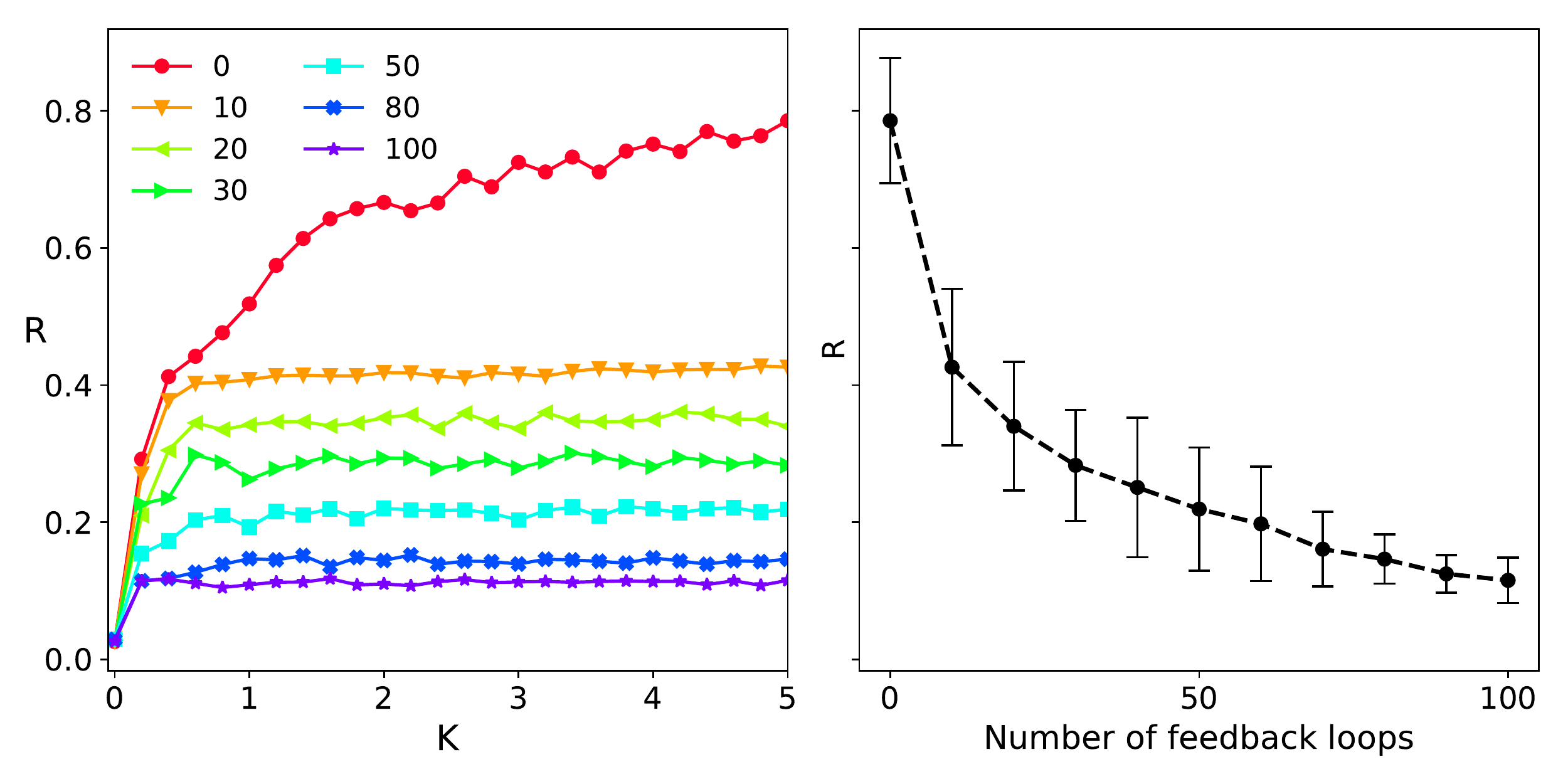}
	\caption{\textbf{Comparison of synchronization of type~I phase oscillators on the scale-free directed networks with different numbers of feedback loops.} The stationary order parameter versus (left)~the coupling strength  and  (right)~the number of feedback loops (for $K=5$). The networks have $N=1000$ nodes and $\gamma=3$. The results are averaged over 30 realizations. }
	\label{feedback}
\end{figure*}

The effect of gradually adding feedback loops to the DAGs is  investigated in the \textbf{figure~\ref{feedback}}. The loops are added randomly to the DAGs and their number is calculated by $\text{Trace}(A^3)/3$. We can see that the synchronizability of the scale-free directed networks decreases nonlinearly with increasing the number of feedback loops. The high standard deviations indicate the fact that in addition to the number of the feedback loops, their positions are also important on the network synchronization.

\subsection*{Merged inhibitory and excitatory feedforward motifs}
\begin{figure*}
\centering
	\includegraphics[scale=0.35, trim=70 10 10 10]{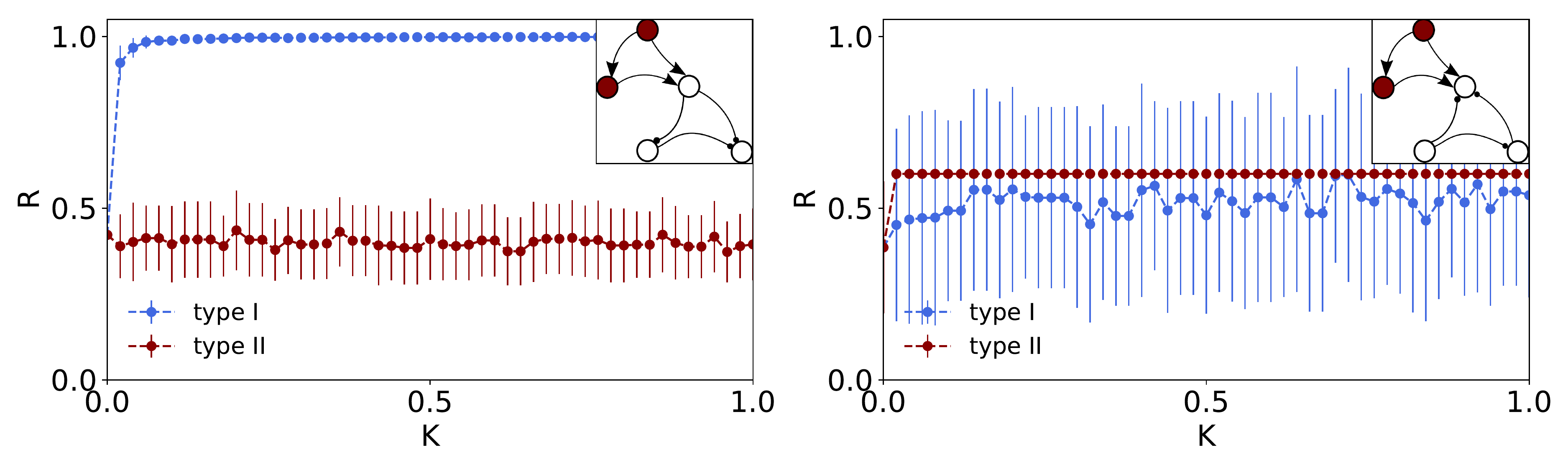}
	\caption{\textbf{Synchronization of the networks constructed from inhibitory and excitatory feedforward motifs.} The stationary order parameters for type~I and type~II phase oscillators versus the coupling strengths for (left)~connected and (right)~disconnected networks. The excitatory and inhibitory oscillators are distinguished by red and white colours. The results are averaged over 30 realizations.} 
	\label{mixed}
\end{figure*}

We show that type~I inhibitory and excitatory connected DAGs (i.e. the networks that have a single source node from which all other nodes are reachable) are synchronized. However, in reality inhibitory and excitatory neurons work together to perform complex tasks. To address this issue, we consider the aggregation of two feedforward motifs that  their effective nodes (i.e. nodes which have outgoing edges) are purely inhibitory or excitatory (see \textbf{Figure~\ref{mixed}}). The results show that, type~I oscillators (but not the type~II ones) are synchronized when these motifs are merged together and form a larger connected network. On the other hand, when the motifs joined to each other and generate a disconnected network, the global synchronization of the oscillators breaks, as one would expect.
The results we derive are not changed qualitatively by swapping excitatory and inhibitory nodes.   In general, excitatory and inhibitory inputs of a neuron are said to be balanced, and this balance is important for the highly irregular firing observed in the cortex. It seems that, disconnected networks constructed from motifs with pure effective nodes can provide these kinds of patterns. Note that, the global dynamics of motifs with hybrid effective nodes is more complex and directly related to the initial phase values. Therefore, further investigations are required to understand the relationship between structure and dynamics of large hybrid networks. 

\section*{Neuron model} 
\subsection*{Characteristics of the neuron models} 

\begin{figure*}
\centering
	\includegraphics[scale=0.4]{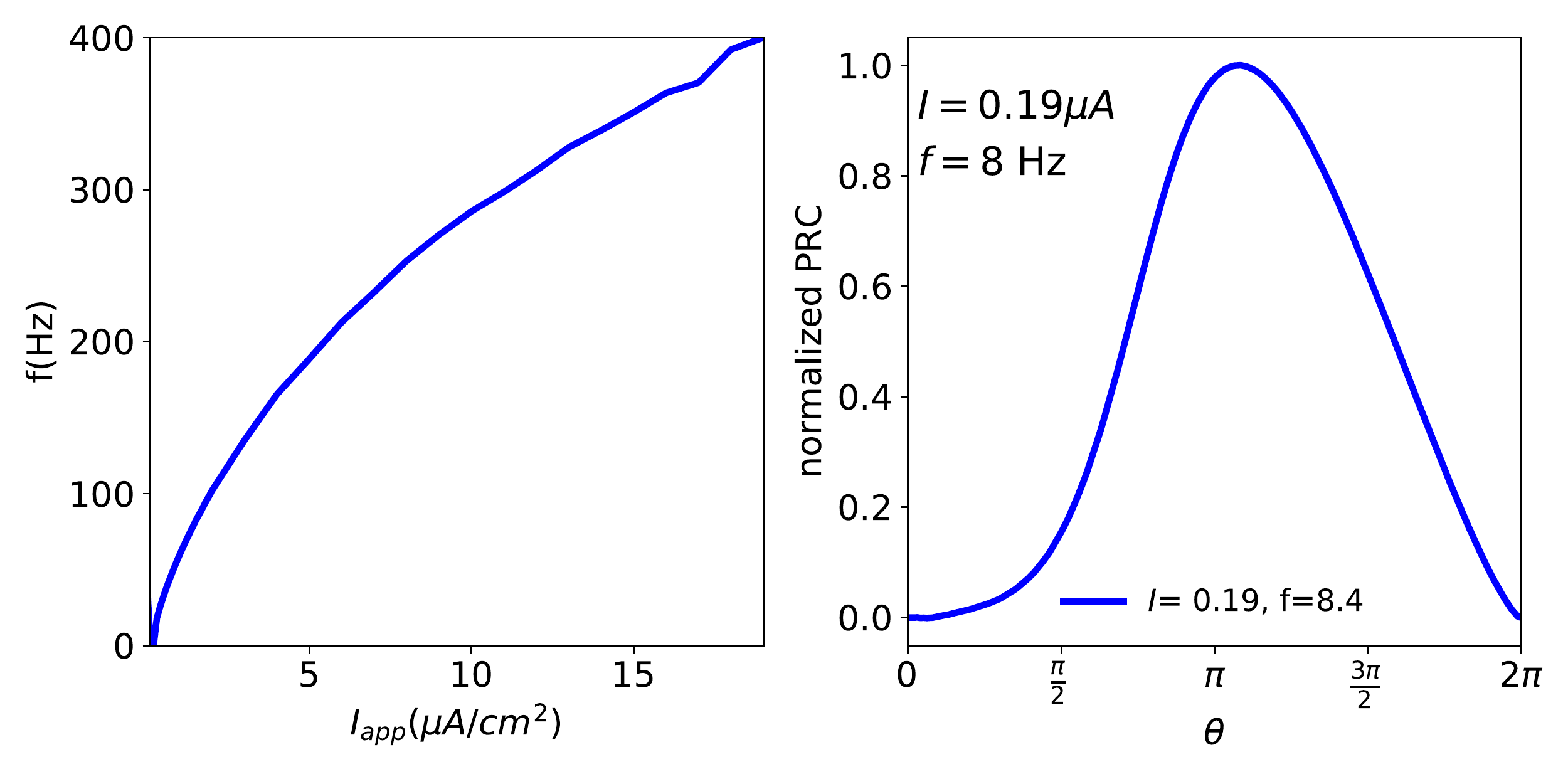}
	\caption{\textbf{Dynamical properties of a type~I Wang-Buzs{\'a}ki inhibitory neuron.} (Left) The firing frequency versus applied current.  (Right) The phase response curve (PRC) of the  neuron.} 
	\label{prci}
\end{figure*}

\textbf{Tables~S1} and \textbf{S2} summarize the values we used in our simulations for the parameters of the Wang-Buzs{\'a}ki and Traub models.
The characteristics of Wang-Buzs{\'a}ki and Traub type~I neurons are presented in the \textbf{figure~\ref{prci}} and \textbf{figure~\ref{prce}}, respectively. As we expected, the frequency-current curve for type~I neurons is continuous and the neuron fires at arbitrarily low frequencies. In addition, the  phase response curve is an exclusively positive curve.

\subsection*{ Voltage synchrony measure}
We also used  the voltage synchrony measure to monitor the degree of spike synchrony in the networks. This measure provides evaluations of long-term fluctuations in the global potential as described by the following formula:
\begin{equation*}
M=\frac{\sqrt{\langle V_g(t)^2\rangle_t-\langle V_g(t)\rangle_t^2 }}{\frac{1}{N}\sum_{i=1}^N \sqrt{ \langle V_i(t)^2\rangle_t-\langle V_i(t)\rangle_t^2}}, \hspace{0.3cm}V_g(t) = \frac{1}{N} \sum_{i=1}^N V_i(t).\\
\end{equation*}
where $N$ is the number of neurons in the network, and $V_g(t)$ is the average membrane potentials of the neurons in the population (global potential).  Here, $\langle\dots\rangle_t$ denotes time-averaging over a large time interval. To normalize the measure, this value is divided by the average fluctuations in the membrane potentials of single neurons. The value of $M$ is bounded between 0 and 1, where $M=1$ and $M=0$ indicate  fully synchronized and asynchronous states, respectively.

The synchronizations of type~I  Wang-Buzs{\'a}ki inhibitory and Traub excitatory neurons connected by scale-free DAGs have been also investigated using voltage synchrony measure (see \textbf{figure~\ref{vneuron1}} and \textbf{figure~\ref{vneuron2}}). The results are similar to that of found using interspike distance synchrony measure (see \textbf{ figures~6} and \textbf{7} in the manuscript). It means that there exist  parameter regions in which type~I inhibitory and excitatory neurons are synchronized on scale-free DAGs.

\subsection*{The effects of parameter $\phi$ of the  Wang-Buzs{\'a}ki model }
\begin{figure*}
	\centering
	\includegraphics[scale=0.32, trim=120 10 10 10]{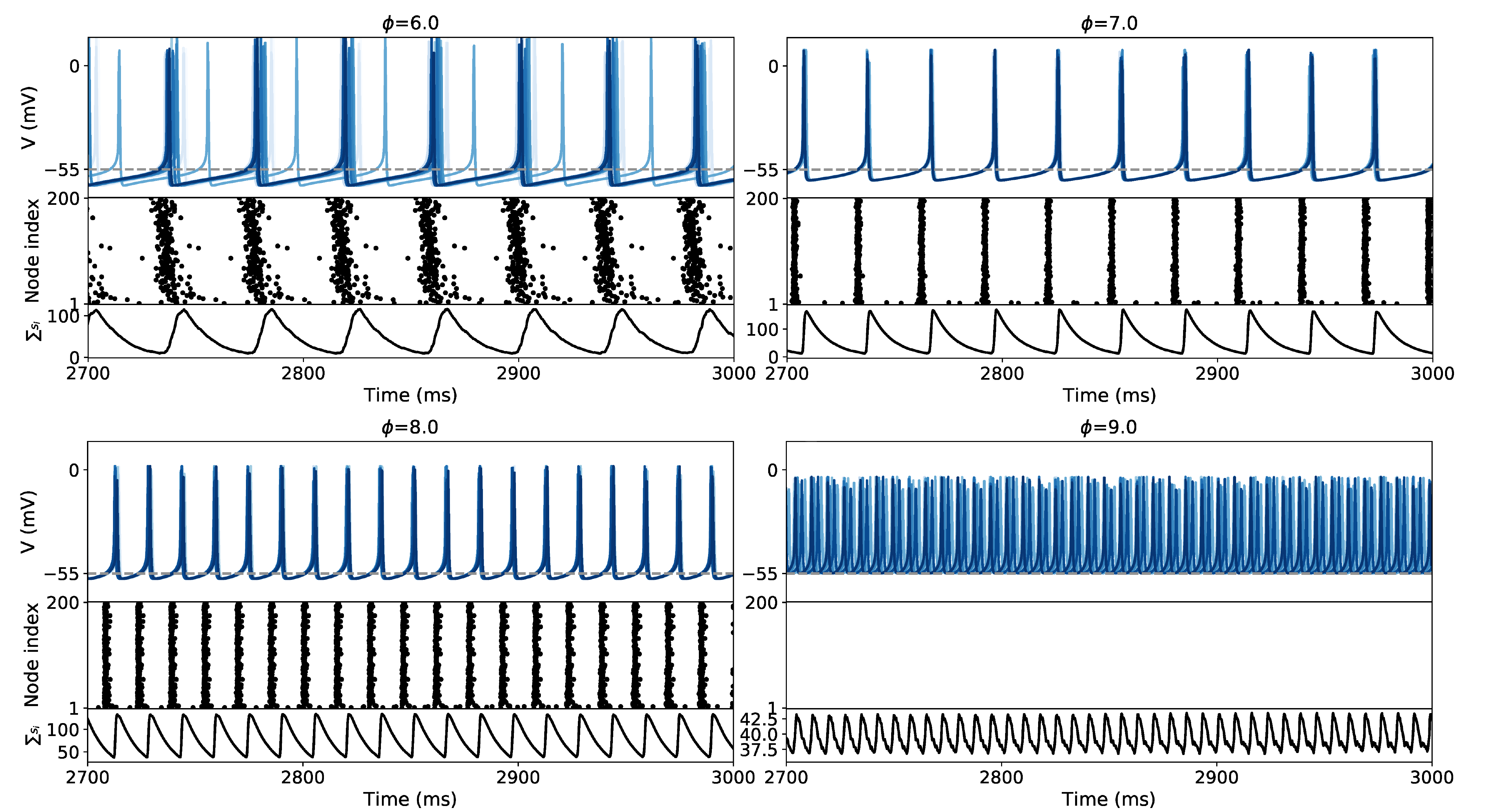}
	\caption{\textbf{Effects of changing $\phi$ on after-hyperpolarization and synchronization of Wang-Buzs{\'a}ki neurons.} From top to bottom, each panel includes the  membrane potential of the neurons  versus time, the raster plot, and the sum of the synaptic gates $\sum_i s_i$ versus time for an specified $\phi$. The networks are scale-free DAGs with $N=200$ nodes and $\gamma=3$. The coupling constant is $K=1.25$.}
\label{phi}
\end{figure*}

\begin{figure*}
	\centering
	\includegraphics[scale=0.4]{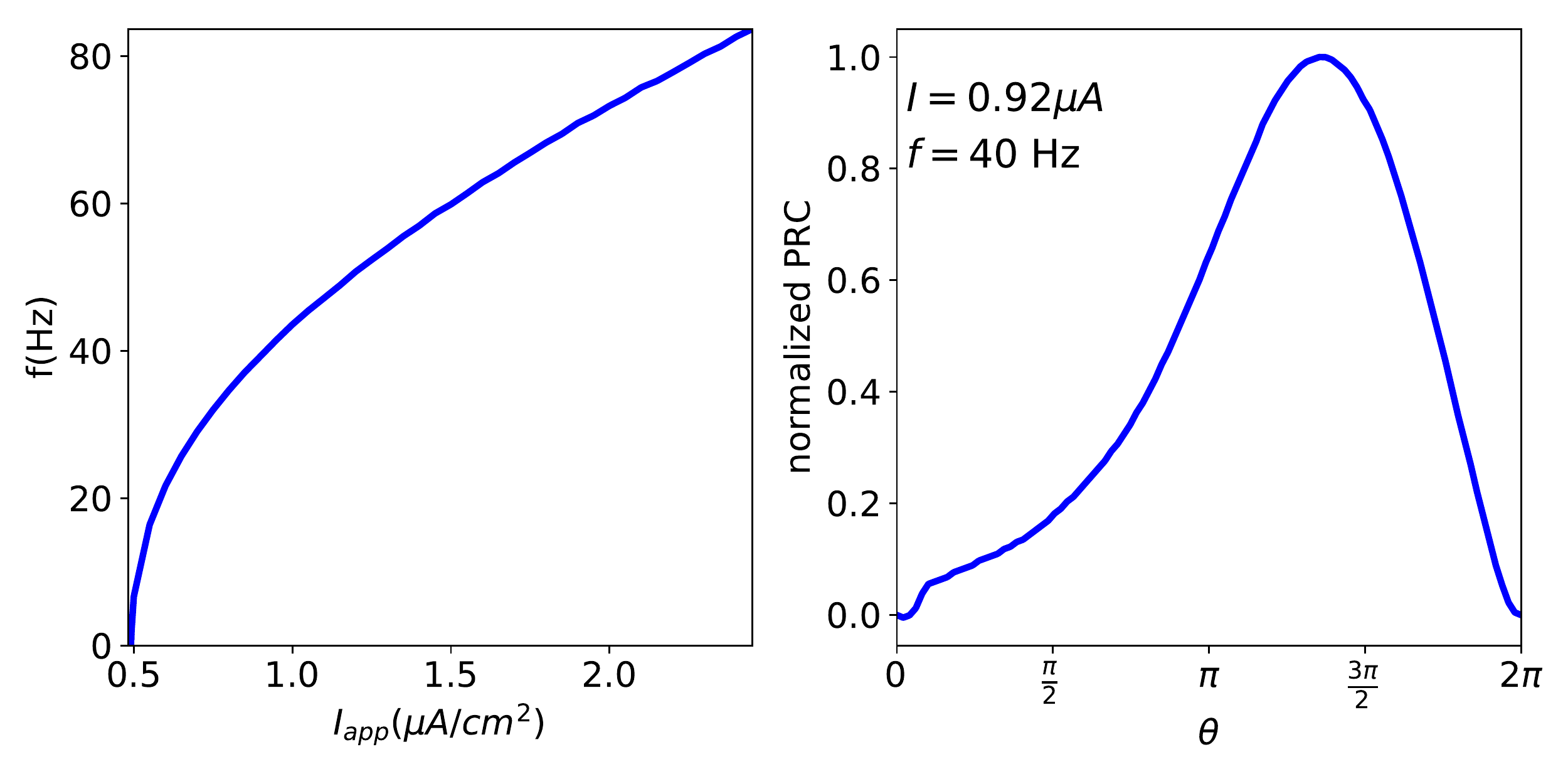}
	\caption{\textbf{Dynamical properties of a type~I Traub excitatory neuron.} (Left) The firing frequency versus applied current.  (Right) The phase response curve (PRC) of the  neuron.} 
\label{prce}
\end{figure*}

\begin{figure*}
	\centering
	\includegraphics[scale=0.32, trim=230 10 10 10]{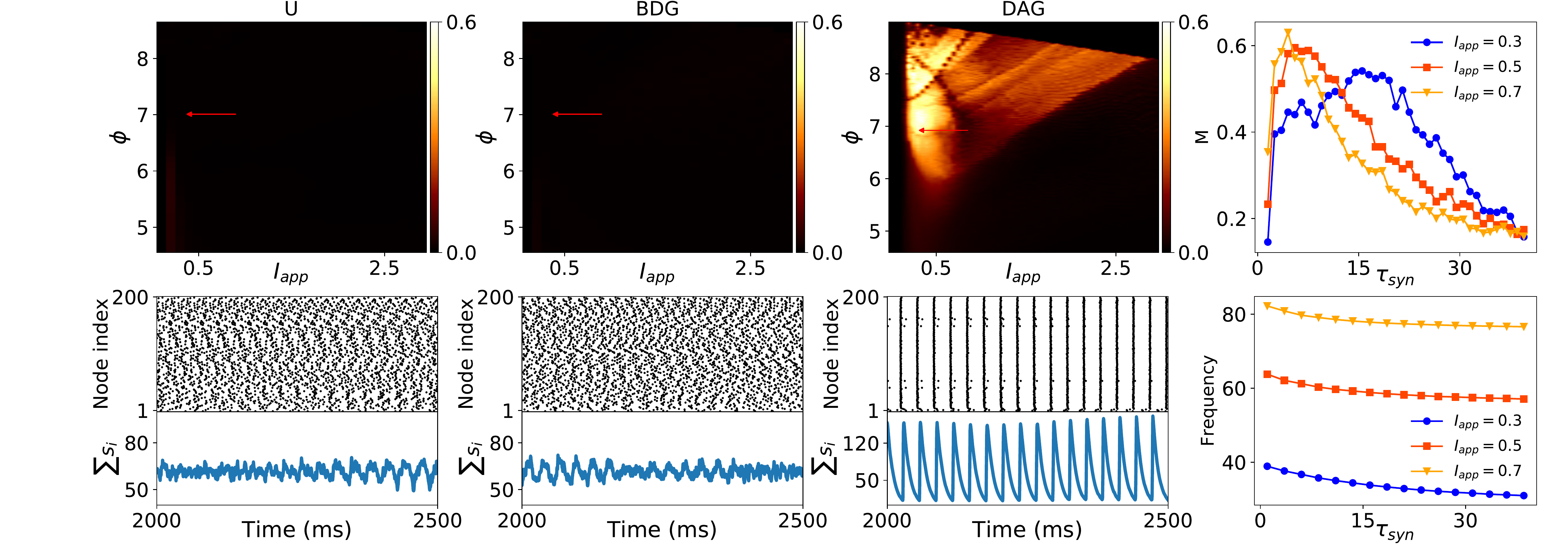}
	\caption{\textbf{Synchronization of Wang-Buzs{\'a}ki inhibitory neurons coupled via scale-free undirected and oriented graphs.}  (First three left columns)~The top panel displays the voltage synchrony of the networks on $I_{app}-\phi$ phase space. The bottom panel represents the raster plots and the sum of the synaptic gates $\sum_i s_i$  for three network instances using the parameters specified by the arrows at the top plots. The synapse parameters are $\alpha=12  ms^{-1}, \beta=0.1 ms^{-1}, g_{syn}=0.1 mS/cm^2$. (Right column)~The top and bottom panels show the effects of synaptic time constant ($\tau_{syn} = 1/\beta$) on synchrony and frequency of oscillation for different values of applied currents, respectively at $K=1.25$. Structural parameters of the networks are valued as $N=200$, and $\gamma=3$.}
\label{vneuron1}
\end{figure*}

\begin{figure*}
	\centering
	\includegraphics[scale=0.32, trim=230 10 10 10]{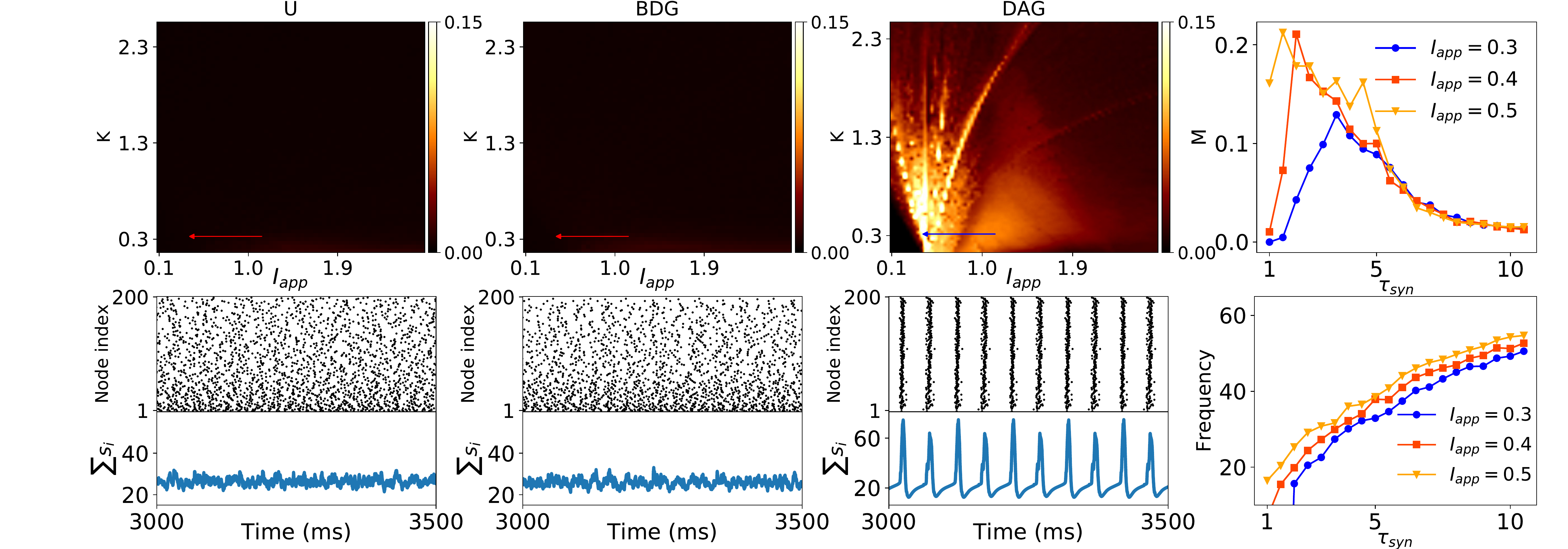}
	\caption{\textbf{Synchronization of Traub excitatory neurons coupled via scale-free undirected and oriented graphs.} (First three left columns)~The top panel shows the voltage synchrony of the networks on $I_{app}-K$ phase space. The bottom panel shows the raster plots and the sum of the synaptic gates $\sum_i s_i$  for three network instances using the parameters specified by the arrows at the top plots. The synapse parameters are $\alpha=12  ms^{-1}, \beta=0.5 ms^{-1}, g_{syn}=0.01 mS/cm^2$. (Right column)~The top and bottom panels represent the effects of synaptic time constant ($\tau_{syn}=1/\beta$) on spike synchrony and frequency of oscillation, respectively at $K=0.3$.  Structural parameters of the network are valued as $N=200$, and $\gamma=3$.}
\label{vneuron2}
\end{figure*}

The parameter $\phi$ in the  Wang-Buzs{\'a}ki mode affects Neuronal afterhyperpolarizations. Decreasing $\phi$  means
reducing $\alpha_h$, $\beta_h$, $\alpha_n$, and $\beta_n$, that is to say slowing down $h$ and $n$. Therefore,  decreasing $\phi$ reduces the firing rate by producing a deep afterhyperpolarization. 
The difference between the lowest value of the voltage and the firing threshold is about 15 mV at  $\phi=5$. This is in agreement with experimental results for fast-firing inhibitory interneurons. However,  for $\phi=9$ this difference disappears and the spikes can not be distinguished (see \textbf{figure~\ref{phi}}).

\begin{table}[]
	\caption{Parameters for Wang-Buzs{\'a}ki neurons.}
	\label{table_wang}
	\centering
	\begin{tabular}{ll}
		\hline	
		parameter			& value 				\\
		\hline
		$g_L$				&  	0.1	(mS/cm$^2$)		\\
		$g_{Na}$			&  	35	(mS/cm$^2$)		\\
		$g_{K}$				&  	9	(mS/cm$^2$)		\\
		$E_L$				&  	-65 (mV)			\\
		$E_{Na}$			&  	55	(mV)			\\
		$E_{K}$				&  	-90	(mV)			\\
		$E_{syn}$			&  	-75	(mV) 			\\
		$g_{syn}$			&  	0.1	(mS/cm$^2$)   	\\
		$\theta_{syn}$		&  	0	(mV)			\\
		$\alpha$			&  	12	(msec$^{-1}$)	\\
		$\beta$				&  	0.1	(msec$^{-1}$)	\\
		$C_m$				&  1.0  ($\mu$F/cm$^2$)	\\
		threshold(spike)	&  -55 (mV)				\\
		\hline
	\end{tabular}
\end{table}

\begin{table}[]
	\caption{Parameters for Traub neuron model.}
	\label{table_traub}
	\centering
	\begin{tabular}{ll}
		\hline	
		parameter			& value 					\\
		\hline
		$g_L$				&  	0.2	 (mS/cm$^2$)		\\
		$g_{Na}$			&  	100	 (mS/cm$^2$)		\\
		$g_{K}$				&  	80	 (mS/cm$^2$)		\\
		$E_L$				&  	-67  (mV)				\\
		$E_{Na}$			&  	50	 (mV)				\\
		$E_{K}$				&  	-100 (mV)				\\
		$g_{syn}$			&  	0.01 (mS/cm$^2$)   		\\
		$\alpha$			&  	12	 (msec$^{-1}$)		\\
		$\beta$				&  	0.5	 (msec$^{-1}$)		\\
		$C_m$				&   1.0  ($\mu$F/cm$^2$)	\\
		$V_{res}$			&   -95 (mV)				\\
		threshold(spike)	&   -55 (mV)				\\
		\hline
	\end{tabular}
\end{table}




\clearpage
\bibliography{lib.bib}{}
\bibliographystyle{unsrt} 